\pdfoutput=1
\documentclass[final]{cvpr}
\usepackage{IEEEtrantools}
\usepackage{times}
\usepackage{graphicx}
\usepackage{amsmath,amssymb,mathrsfs,latexsym,bbm} 

\usepackage{diagbox}
\usepackage{xcolor,rotating,epic,eepic}
\usepackage{color, colortbl}
\usepackage{soul}

\usepackage[pagebackref=true,breaklinks=true,colorlinks,bookmarks=false]{hyperref}

\newcommand\green[1]{\textcolor{green}{#1}}
\newcommand\magenta[1]{\textcolor{magenta}{#1}}
\newcommand\olive[1]{\textcolor{olive}{#1}}
\newcommand\brown[1]{\textcolor{brown}{#1}}

\newcommand\black[1]{\textcolor{black}{#1}}
\definecolor{Gray}{gray}{0.9}
\definecolor{LightCyan}{rgb}{0.88,1,1}

\def\un{{\mbox{\rm 1\hspace{-0.26em}l}}}
\def\repsu{A}
\def\repsku{f}
\def\repshu{g}

\def\ssbs{u}
\def\softplus{\textrm{softplus}}

\def\mish{PMISH}
\def\misho{MISH}

\def\swish{PSWISH}
\def\swisho{SWISH}

\def\genfbf{{\mathcal E}}

\def\exp{e^}

\def\Rset{\mathbb{R}}

\def\Rset{\mathbb{R}}

\def\un{{\mbox{\rm 1\hspace{-0.26em}l}}}

\DeclareMathOperator{\sgn}{sgn}

\newcommand\blue[1]{\textcolor{blue}{#1}}
\newcommand\red[1]{\textcolor{red}{#1}}

\begin{document}

\title{Parametric Rectified Power Sigmoid Units: Learning Nonlinear Neural Transfer Analytical Forms}

\author{%
{Abdourrahmane M. {\sc ATTO}\thanks{Email: \texttt{Abdourrahmane.Atto@univ-smb.fr} - Phone: +334 50 09 65 27 - Fax: +334 50 09 65 59 }\thanks{
The work was supported by DAR START DEEP APR-6177/5930 grant of the CNES - France.
}}
\\
{  Universit\'e Savoie Mont Blanc - France} 
    \and
  {Sylvie {\sc GALICHET}}
\\
{  Universit\'e Savoie Mont Blanc - France}   
\and
  {Dominique {\sc PASTOR}}
\\
{  TLECOM Bretagne - France} 
\and
  {Nicolas {\sc M\'EGER}}
\\
{  Universit\'e Savoie Mont Blanc - France} 
}


%
\maketitle
%
\begin{abstract}
The paper proposes representation functionals in a dual paradigm where learning jointly concerns both linear convolutional weights and parametric forms of nonlinear activation functions. The nonlinear forms proposed
to perform the functional representation are associated with a new class of parametric neural transfer functions called rectified power sigmoid units. This class is constructed to 
benefit from the advantages of both
sigmoid and rectified linear unit functions, 
while rejecting their respective drawbacks.
Moreover, the analytic form of this new neural class involves scale, shift and shape parameters so as to obtain a wide range of activation shapes, including the standard rectified linear unit as a limit case. Parameters of this neural transfer class are considered as learnable for the sake of discovering the complex shapes that can contribute 
to solving machine learning issues. Performance achieved by the  joint learning of convolutional and rectified power sigmoid learnable parameters are shown outstanding in both shallow and deep learning frameworks. This class opens new prospects with respect to machine learning in the sense that learnable parameters are not only attached to linear transformations, but also to suitable nonlinear operators.
\end{abstract}

%

\section{Introduction}
\label{introduction}

Standard neural transfer functions such as Rectified Linear Unit (ReLU) \cite{relu10} and sigmoid hereafter, denoted respectively $U$ and $S$ with 
\begin{equation}
U(x) = x \un_{x > 0}   = \max\left(0, x \right)
\label{eqrelu}
\end{equation}
\begin{equation}
S(x)= \frac{1}{1+\exp{-x}}
\label{eqsigmoid}
\end{equation}
are non-parametric functions in the sense that their analytic expressions do not depend on unknown parameters or weights. 
While the sigmoid function has been the leader of the early stage neural transfer functions, it has been outclassed by the ReLU in most recent deep Convolutional Neural Networks (CNN), see
\cite{alexnet15,
vgg192014,
resnet101,
googlenet15,
Carreira17,
shi18,
lin20} among others.

In terms of machine learning, the first major difference between ReLU and sigmoid is the fact that ReLU output is expected to be a sparse sequence in general, while sigmoid function simply penalizes its entries without forcing non-zero values to zero. Thus, in terms of the compromise between computational complexity and available working memory, ReLU is naturally favored when very deep networks are under consideration.

The second major difference between ReLU and sigmoid concerns their derivatives.
The derivative $U^\prime$ of $U$ is the Heaviside unit step function: such a function is stable by composition. 
However, it admits a singularity at 0 and has the same constant output for both small and large positive values, which may be counterintuitive since if we consider for instance sparse transforms, small and large positives does not carry the level of information. 
In addition, because of the zero-forcing operated by ReLU derivative, then learning can be inhibited\footnote{Leaky ReLU: 
$
x \mapsto 
x \un_{x > 0} + 0.01 x \un_{x \leqslant 0}
$
can avoid such issues, however, it is less used in deep neural networks because it raises other issues (such as the arbitrary penalization of negative values, the latter being far from bio-inspired behaviors).
} in a ReLU CNN when the processing implies a large amount of negatives.

In contrast with ReLU, the derivative of a standard sigmoid is smooth everywhere. However, it is always strictly less than 1 and this can also lead to a fast decrease to 0 of the sigmoid increments by composition and this, both for positive and negative entries. 

One can note that both ReLU and sigmoid admit parametric forms 
$
x \mapsto 
x \un_{x > 0} +  \alpha x \un_{x \leqslant 0}
$
for parametric ReLU \cite{He2015ICCV}
and 
$x \mapsto  S(\alpha x)$ for parametric sigmoid \cite{sigmoid78}, \cite{sigmoid95}.
These parametric forms can solve the limitations highlighted above for specific applications and when $\alpha$ is chosen carefully. It is worth noticing that the use of these parametric forms is limited to specific datasets or specialized networks and their generalization capabilities need to be proven.

In terms of image processing, important properties are invariances by rotation, translation and scaling. It is well known that rotation invariance can be handled by a suitable sequence of convolution filters. 
For the two remaining invariance properties: on the one hand, both ReLU and sigmoid are translation-variant. On the other hand, only ReLU is scale-invariant, but from a general perspective, translation and scaling invariances can also be obtained from other components of the network such as pooling and convolution layers respectively for the translation and scaling invariances.

This paper provides in the Section \ref{sec repsu}, new neural transfer functions that possess 
most
of the
desirable properties highlighted above, while limiting the undesirable ones. 
Because biological neurons have non-uniform\footnote{
The activation functions depend on the specialization and the depth of the neurons in the brain as diverse inhibition mechanisms in the brain can influence information transfer.
} activation functions, we will propose in Section \ref{sec repsulearn}, a convolutional neural learning framework where learning includes the determination of suitable transfer function with respect to the depth of the layer. Despite the fact that this framework leads to a higher computational complexity than using a non-parametric ReLU transfer 
functions, 
we will show in Section \ref{sec repsulearn} that it is highly relevant for machine learning by providing comparisons with respect to analog frameworks based on non-learnable ReLU, MISH \cite{mish20} or SWISH \cite{swish17} nonlinearities.
Section \ref{sec conclude} concludes the work and provides outlooks raised by the joint linear-and-nonlinear learning framework.


\begin{figure}[!t]
\centering
\hspace*{-0.45cm}
\begin{tabular}{@{}c@{}}
 \includegraphics[height=8.00cm]{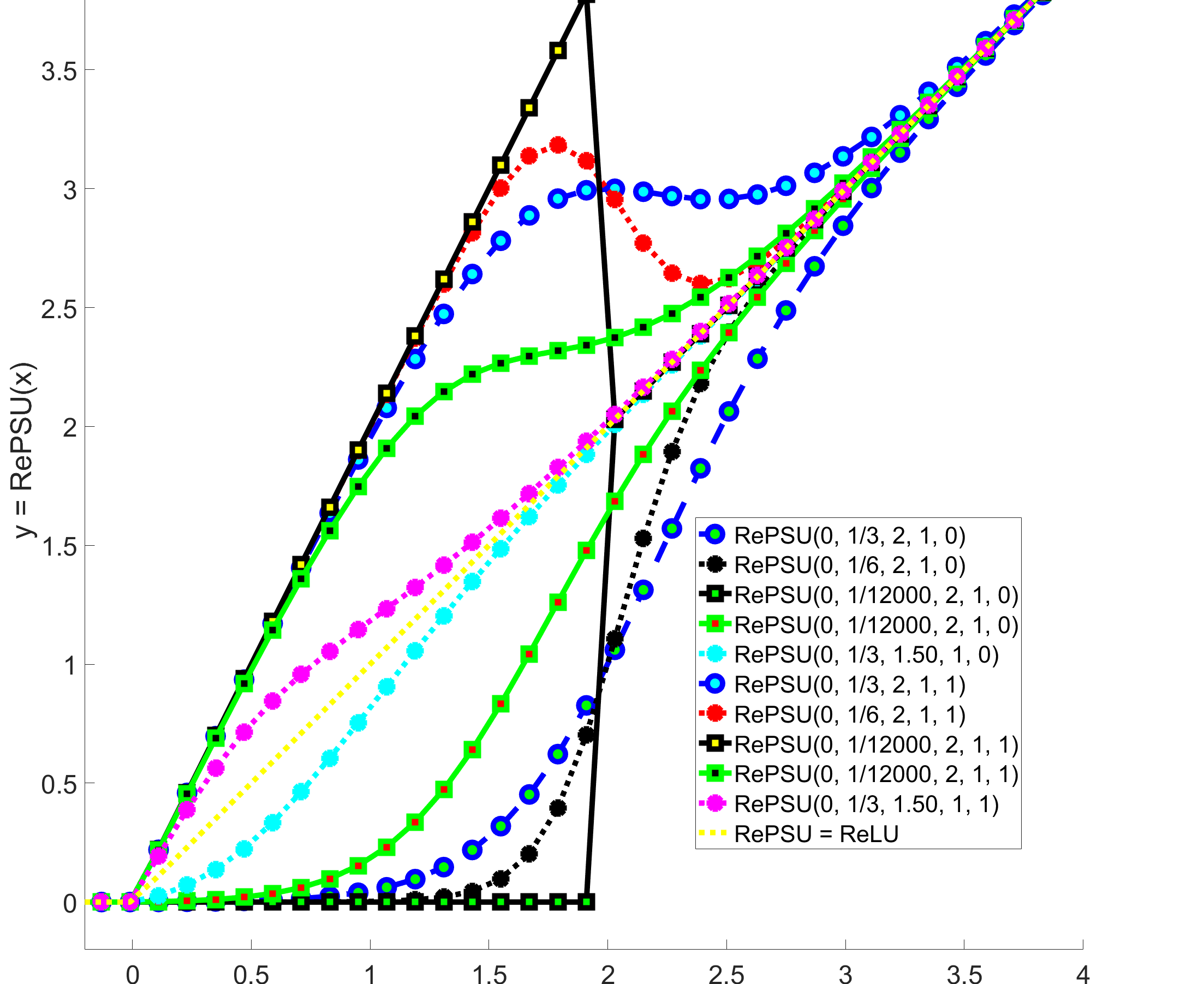}
 \\
 \includegraphics[height=8.00cm]{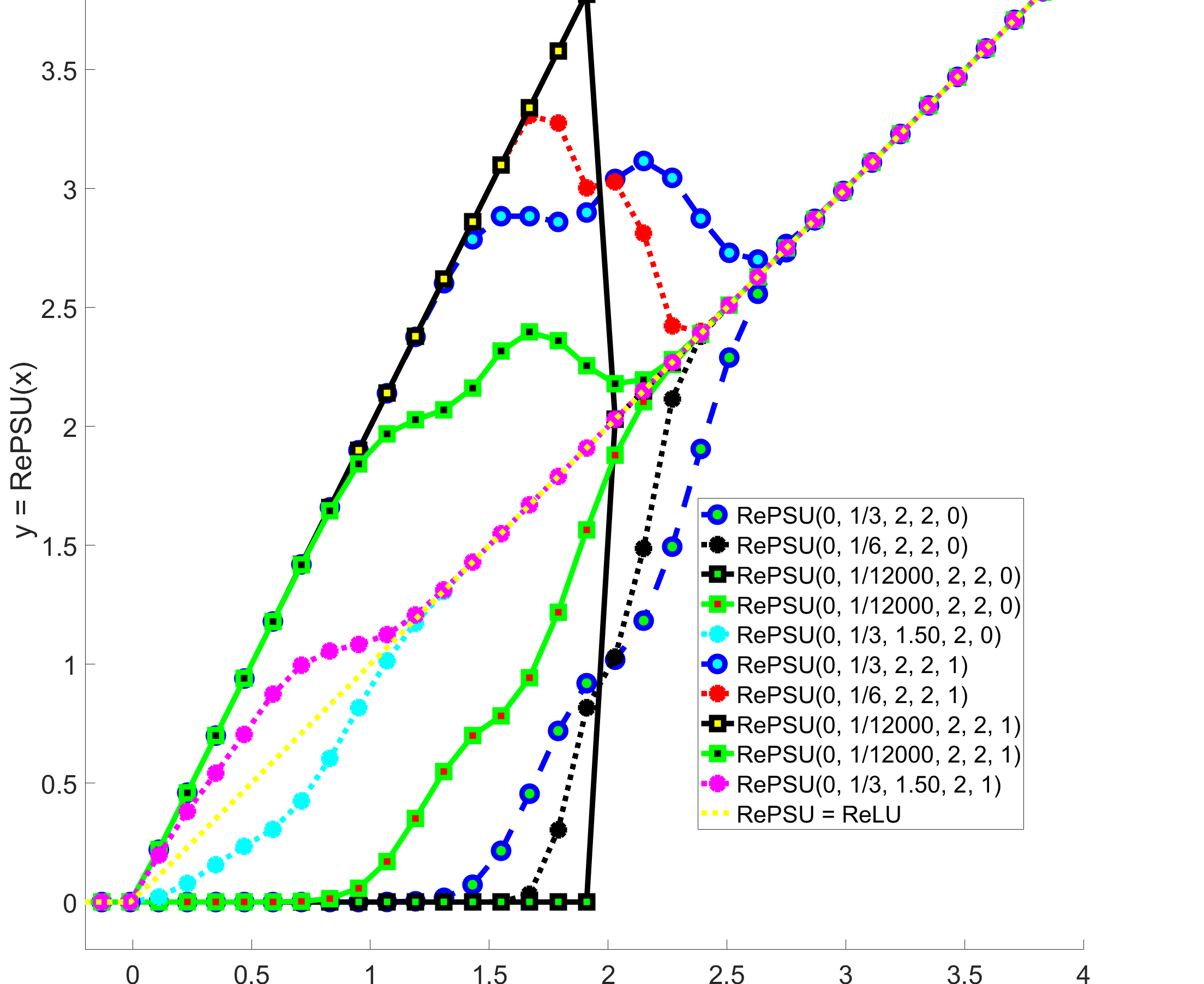}
\end{tabular}
\caption{Graphs $y = \repsu_{\lambda, \sigma, \mu, \beta, \alpha }(x)$ of RePSU for different parameters $\sigma, \beta, \alpha$ when $\lambda=0$ and $\mu=2$.
Intuition: large positives must be forwarded quasi-unchanged to the upstream part of  the neural network. Small positives can either be attenuated (case of RePSKU) because they do not carry enough information (noise), or be amplified (case corresponding to RePSHU) because they are associated with weak signals carrying significant information. Negative values (smaller than $\lambda=0$) are forced to zero.
These behaviors are biologically plausible.
}
\label{figRePSU}
\end{figure}

\section{Rectified power sigmoid shrinking and stretching units}\label{sec repsu}

Let 
$\sgn$ denotes the sign function and 
$\un_\mathcal{E}$ be the standard indicator function of set $\mathcal{E}$. We will use the notation:
\begin{equation}
\un_{ \lambda  } (x) = 
\un_{ \{ x \geqslant \lambda \} }= 
\left\{\displaystyle
\begin{array}{ccc}
1 & \textrm{if} & x \geqslant \lambda \\
0 & \textrm{if} & x < \lambda \\
\end{array}
\right.
\label{eqindic}
\end{equation}
We define the \blue{Rectified Power Sigmoid shrinKage Units (RePSKU)} by the parameterized form:
\begin{equation}
\repsku_{\lambda, \sigma, \mu, \beta}(x)= 
\displaystyle
\frac{
(x - \lambda) \un_{ \lambda  } (x) 
}{
1 + \exp{-   \sgn(x-\mu)    \left( \frac{\vert x-\mu \vert}{\sigma} \right)^\beta    }
} 
\label{eqRePSKU}
\end{equation}
and the \blue{Rectified Power Sigmoid stretcHage Units (RePSHU)} as:
\begin{equation}
\repshu_{\lambda, \sigma, \mu, \beta}(x)
=
2x \un_{ \lambda  } (x) 
-
\repsku_{\lambda, \sigma, \mu, \beta}(x)
\label{eqRePSHU}
\end{equation}

The main contribution provided by the paper is the so-called \blue{Rectified Power Sigmoid Unit (RePSU)} activation class defined by the integration of RePSKU and RePSHU in the following parametric form: 
\begin{equation}
\repsu_{\lambda, \sigma, \mu, \beta, \alpha }(x)
=
\alpha \repshu_{\lambda, \sigma, \mu, \beta}(x)
+
(1-\alpha)
\repsku_{\lambda, \sigma, \mu, \beta}(x)
\label{eqRePSU}
\end{equation}

REPSU \emph{threshold} $\lambda$ is inspired from the behavior of ReLU functions (see by Eq. \eqref{eqrelu} in particular for $\lambda = 0$ which implies forcing negative inputs to 0).
REPSU involves in its exponential term,
a \emph{shift} parameter $\mu$, a \emph{scale} parameter $\sigma$ and a \emph{shape} parameter $\beta$: these parameters are inspired from the generalized Gaussian distribution, but integrated as in \cite{attoICASSP08} (Smooth Sigmoid Based Shrinkage functions, SSBS) by using a sigmoid form so that REPSU corresponds effectively to an activation class. 


Examples of RePSU graphs are given by Figure \ref{figRePSU} for different parameters $(\sigma, \beta, \alpha)$ considered hereafter as positive real values. Graphs corresponding to RePSKU (respectively RePSHU) are located under (respectively over) the diagonal representing $y=x$.

\medskip
\noindent
\textbf{Remark}

If $\alpha=0$ and $\beta = 1$, then function $\ssbs_{\lambda, \xi , \mu} =  \repsu_{\lambda, 1/\xi, \mu, 1, 0 }$ have the following form:
\begin{equation}
\ssbs_{\lambda, \xi, \mu}(x)= 
\left\{\displaystyle
\begin{array}{ccc}
\frac{x - \lambda}{
1+\exp{-\xi(x-\mu)}
} 
& \textrm{if} & x \geqslant \lambda \\
0 & \textrm{if} & x < \lambda \\
\end{array}
\right.
\label{eqReSKU}
\end{equation}
where we have assumed $\xi = 1/\sigma$.
Function $\ssbs_{\lambda, \xi , \mu}$ corresponds to the restriction on $\Rset^+$ of the SSBS activation functions \cite{attoICASSP08}.
Furthermore, 
the restriction on $\Rset^+$  of the SSBS class includes the SWISH
 \cite{swish17} (when $\mu=0$) and SiLU \cite{silu20} (for $\mu=0$ and $\xi=1$) activations\footnote{The restriction of SWISH and SiLU on $\Rset^-$ is composed by negligible values that are not forced to zero: a limitation in terms of sparsity that is avoided by the ReLU-like behavior of $\ssbs_{0, \xi, \mu}(x)$ on $\Rset^-$.}.
 Moreover, RePSU class includes standard ReLU since its subclass
$\ssbs_{\lambda \rightarrow 0, \xi \rightarrow +\infty,\mu} = U$
where $U$ is the ReLU function defined by Eq. \eqref{eqrelu}.
 Thus $\repsu_{\lambda, \sigma, \mu, \beta, \alpha }$ can be seen as a generalization of several standard transfer functions.



\begin{figure*}[!t]
\centering
\begin{tabular}{lr}\hline\hline
\multicolumn{2}{c}{ReSKU functions with respect to parameter variation}\\
 \includegraphics[height=4.50cm]{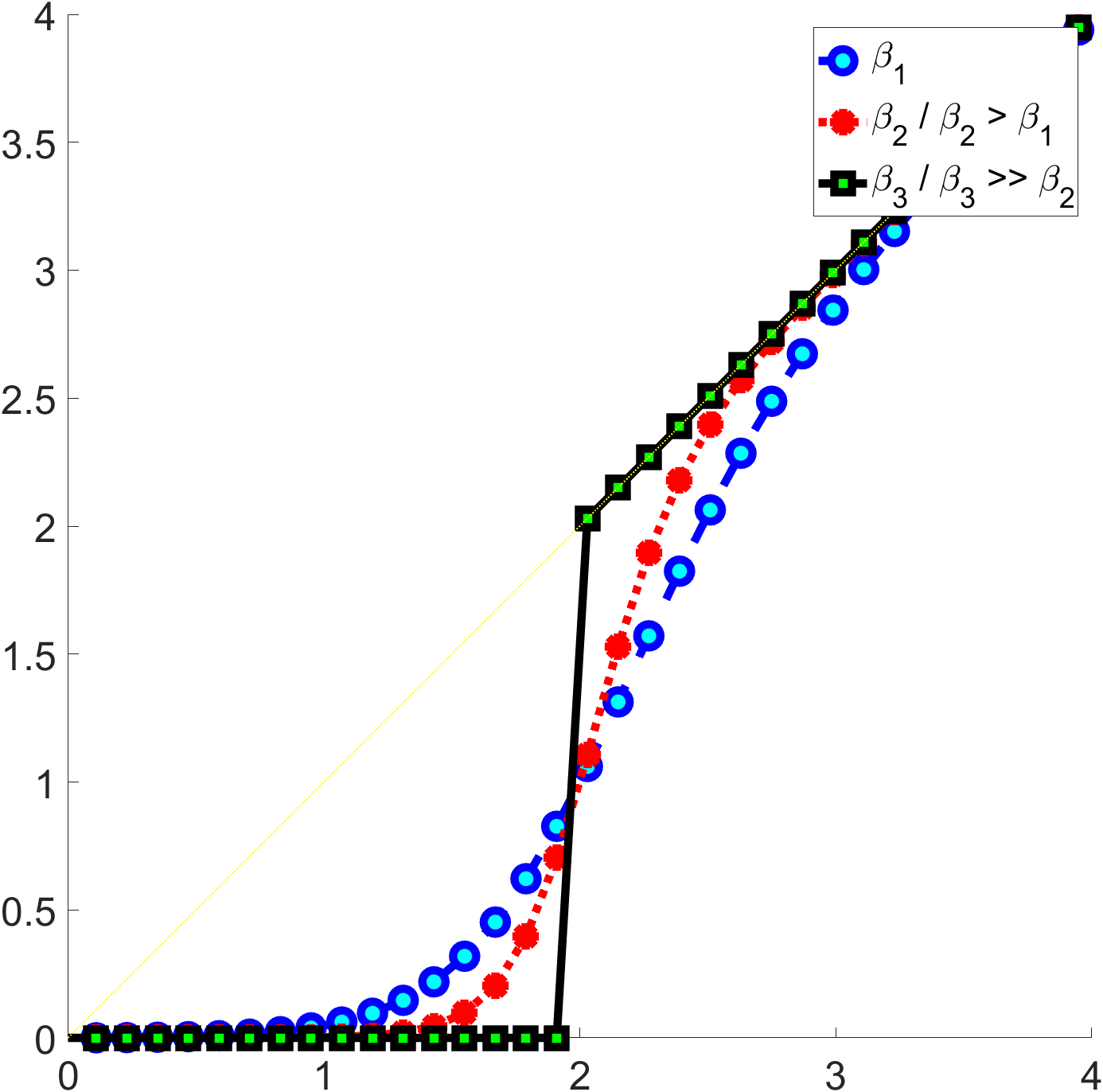}
& 
  \includegraphics[height=4.50cm]{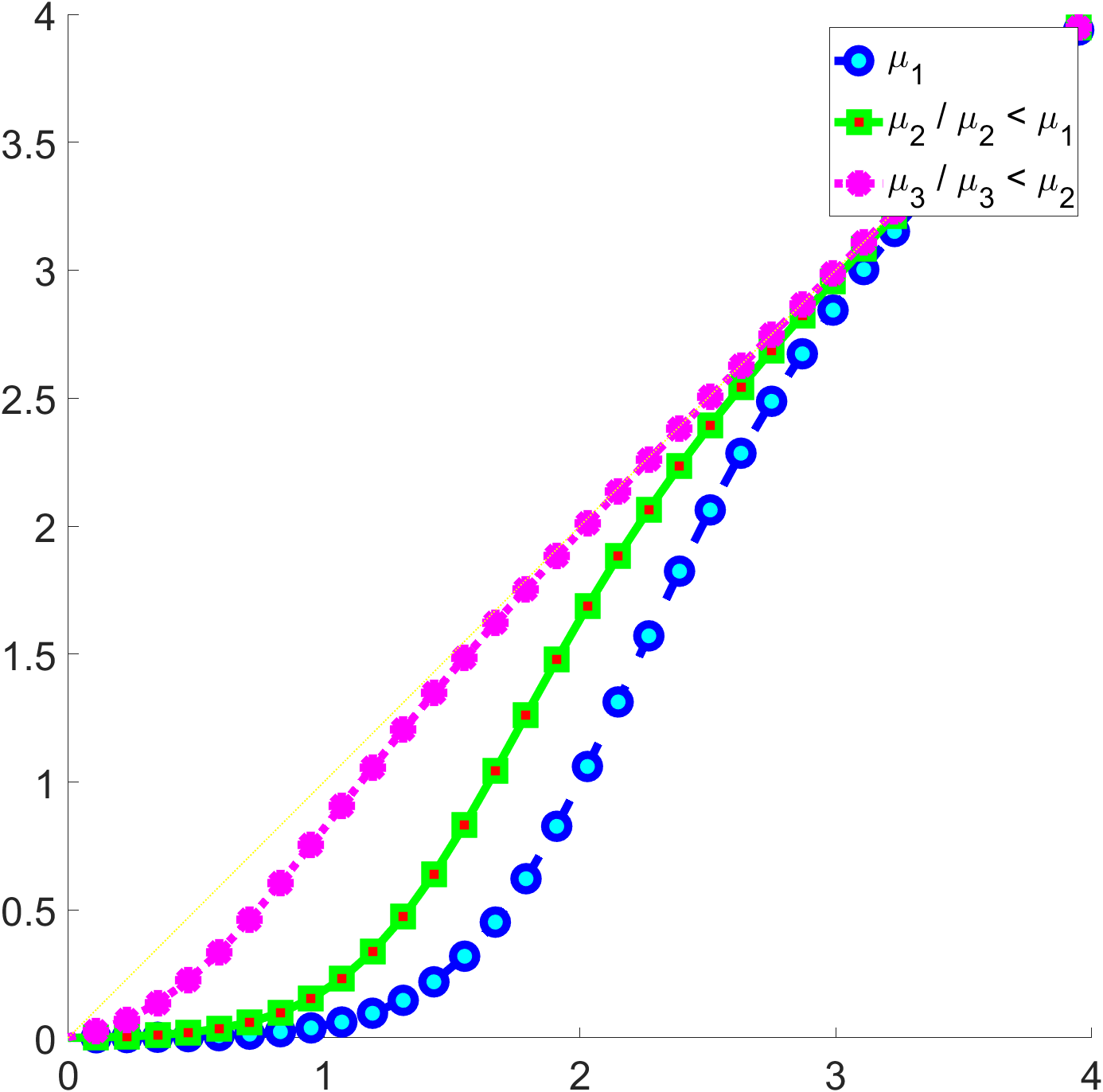} 
   \\\hline\hline
\end{tabular}
\caption{Examples of ReSKU $\ssbs_{\lambda, \xi, \mu}$ shapes depending on parameters $\xi$ and $\mu$, with $\lambda =0$. We recall that $\ssbs_{\lambda, \xi, \mu}(x) = 0$ for $x<\lambda$. Left: $\mu =2$. Right: $\xi = 3$.}
\label{figReSKU}
\end{figure*}

\begin{figure*}[!t]
\centering
\begin{tabular}{lr}\hline\hline
   \multicolumn{2}{c}{ReSKU derivatives with respect to parameter variation}\\\hline
   \includegraphics[height=4.50cm]{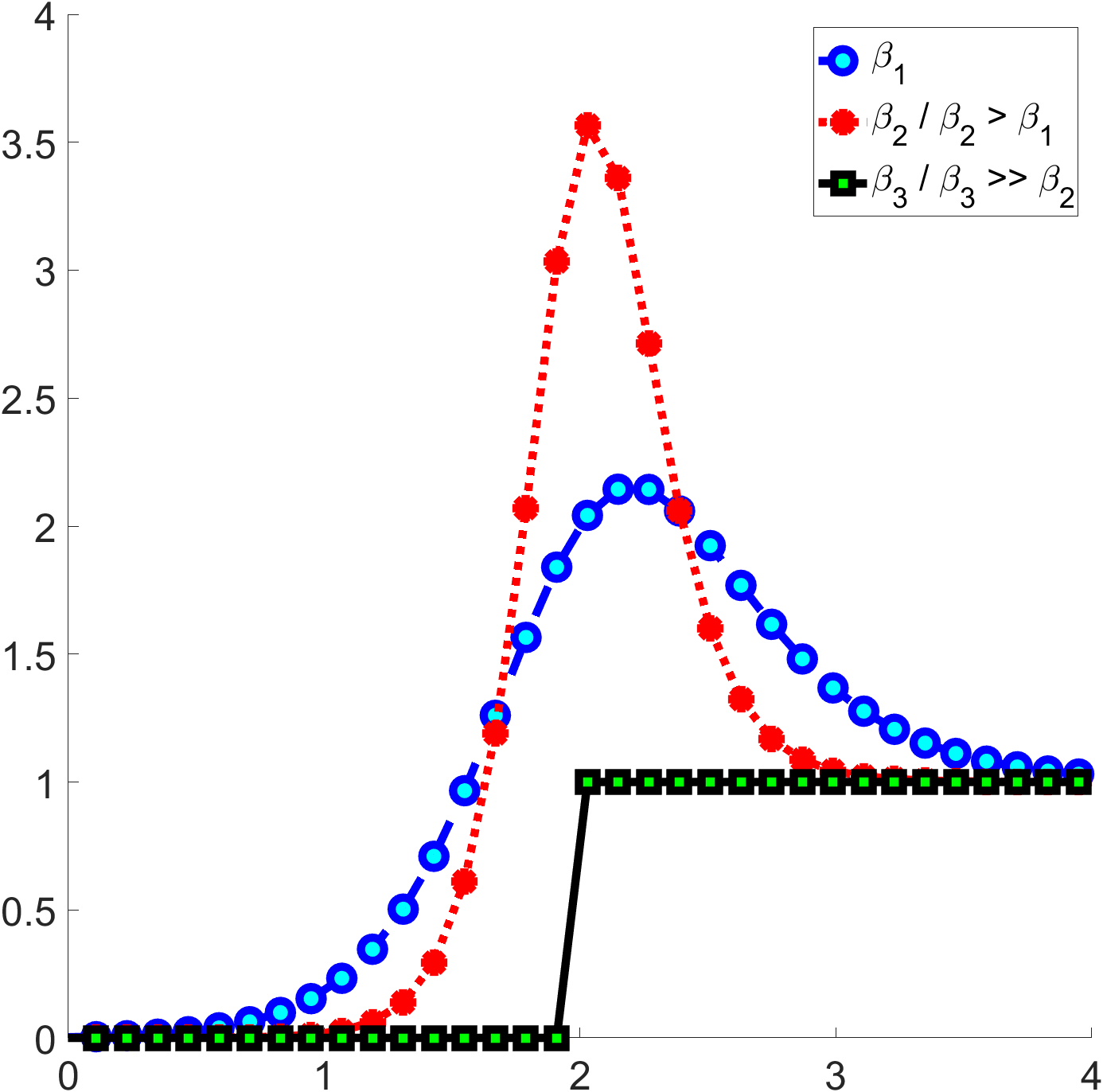}
& 
  \includegraphics[height=4.50cm]{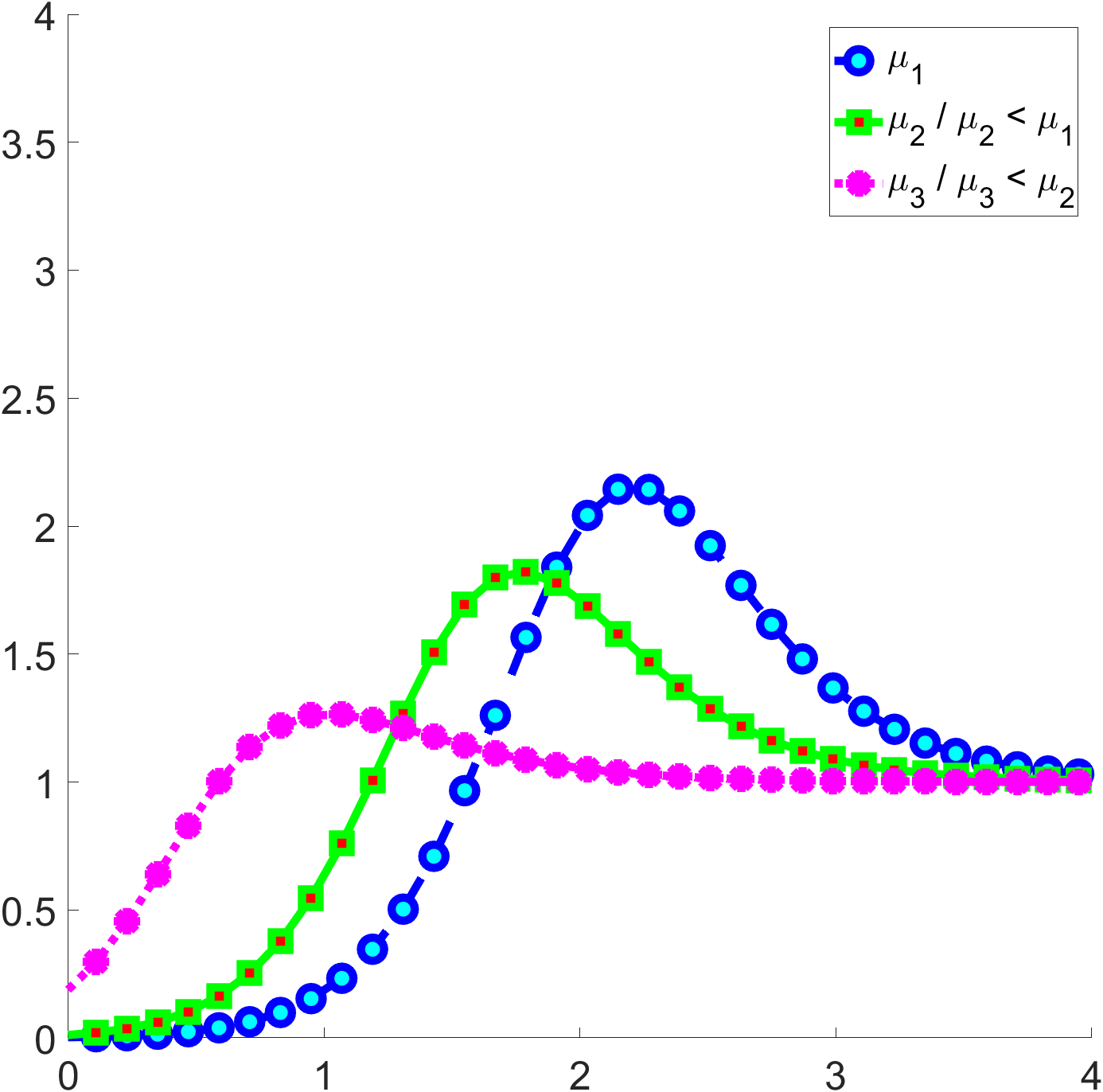}  \\\hline\hline
\end{tabular}
\caption{Derivatives of the ReSKU functions given in Figure \ref{figReSKU}. }
\label{figReSKUprime}
\end{figure*}

The following provides additional properties of ReSPU class. We will focus on the shrinkage subclass with fixed shape parameter $\ssbs_{\lambda, \xi, \mu}$ defined by Eq. \eqref{eqReSKU} and this, for the sake of conciseness since all properties cannot be summarized in a short size paper.
Since $\ssbs_{\lambda, \xi, \mu}$ does not involve the power of the shape parameter $\beta$, then $\ssbs_{\lambda, \xi, \mu}$ will be simply called the class of Rectified Sigmoid shrinKage Units (ReSKU).

\subsection{Intra-class translation invariance for ReSKU}\label{ssecttranslate}

The invariance highlighted below applies at ReSKU subclass level. Assumes $x-\tau$ is the input of the ReSKU class. Then one can note that for ReSKU functions defined by Eq. \eqref{eqReSKU}, we have:
$$
\ssbs_{\lambda, \xi, \mu}(x-\tau) = \ssbs_{\tau+\lambda, \xi, \tau+\mu}(x) 
$$
This implies that the output of a shifted input can be deducted directly from parameter shifts of the ReSKU. Thus, translation invariance can be achieved by a series of ReSKU functions associated with different parameters.   
In comparison with ReLU for which translation can induce forcing to zero, ReSKU has the capability to either keep invariant or force to zero a given value depending on the training objective. 

\subsection{Intra-class scaling conservatives for ReSKU}\label{ssecscaling}

Scaling is present at different stages in image processing. For instance, dividing an 8-bit coded image by a positive constant changes the scaling, but does not affect pixel distribution shapes. For ReSKU functions given by Eq. \eqref{eqReSKU}, we have the following property:
$$
\ssbs_{\lambda, \xi, \mu}(\alpha x) = \alpha \ssbs_{\lambda/\alpha, \alpha \xi, \mu/\alpha}(x) 
$$
Thus, re-scaling a value can be inferred by re-scaling ReSKU outputs thanks to scaled parameters set as $(\lambda^\prime, \xi^\prime, \mu^\prime) = (\lambda/\alpha, \alpha \xi, \mu/\alpha)$.

\subsection{Derivative properties for ReSKU}\label{ssecderivative}


From Eq. \eqref{eqReSKU}, the derivatives of ReSKU are 
\begin{equation}
\begin{array}{@{~}r@{~}c@{~}l@{~}}
\ssbs_{\lambda, \xi, \mu}^\prime(x) &=&
\left\{\displaystyle
\begin{array}{ccc}
\frac{ 1+ 
\xi \ssbs_{\lambda, \xi, \mu}(x) \exp{-\xi(x-\mu)} }{
1+\exp{-\xi(x-\mu)}
} 
& \textrm{if} & x > \lambda \\
0 & \textrm{if} & x < \lambda \\
\end{array}
\right.
\label{eqsigmoidder}
\\
\\
&=&
\left\{\displaystyle
\begin{array}{ccc}
\xi \ssbs_{\lambda, \xi, \mu}(x)
+
\frac{1-\xi \ssbs_{\lambda, \xi, \mu}(x) }{
1+\exp{-\xi(x-\mu)}
} 
& \textrm{if} & x > \lambda \\
0 & \textrm{if} & x < \lambda \\
\end{array}
\right.
\end{array}
\end{equation}
Examples of ReSKU functions are given in Figure \ref{figReSKU} and Figure \ref{figReSKUprime} highlights smoothness of the derivatives associated with these functions: the general behavior is a ``no-jump'' property\footnote{
The derivative has no-jump, expected for limit parameters.
} which implies introducing less va\-ria\-bi\-li\-ty\footnote{
Discontinuities are known to generate a high variance in iterative processing
}
 between close objective values. 
In contrast, at the limit corresponding to translated versions of ReLU  (when $\xi \rightarrow +\infty$ and for fixed $\mu$), the derivative shifts from 0 to 1 in passing from 0.
The outstanding ReSKU property is that the graph of the derivative can be flat if desired (case for the convergence to a standard ReLU). In addition, all ReSKU derivatives asymptotically tend to 1 at infinity: the ReSKU behavior is very stable for very large input values.

The following addresses performance of CNN involving ReSKU nonlinearities.


\section{Learning both linearities and nonlinearities}\label{sec repsulearn}

The second main contribution provided by the paper is the joint learning of standard convolutional linearities and parameters of the non-linear ReSPU class: ReSPU parameters $\lambda, \sigma, \mu, \beta, \alpha$ (see Eq. \eqref{eqRePSU}) are assumed learnable hereafter.
The issue addressed in this section is then measuring the performance adduced in learning ReSPU nonlinearities in a CNN, in comparison with the alternative used in standard ReLU-CNN approaches based on learning only linear parameters. 
For the sake of limiting computational complexity, only a single  ReSPU layer will be considered in the CNN whatever the deepness of the latter: additional nonlinearities will be composed by ReLU in order to go deeper without increasing significantly the computational complexity of the framework.

We recall that the main goal of the paper is to prove the interest in learning optimal nonlinear activations from a family of functions including the standard ReLU (in contrast with using directly the standard non-learnable ReLU).
So, a good comparison should address a ``learnable ReSKU architecture'' \emph{versus} a ``purely ReLU architecture''.

However, we extend (see Tables \ref{tab cnnbasic}, \ref{tab cnndeep}, \ref{tab montecarloRePSU} and \ref{tab gfbfcnnperfo}) the comparison by testing both fixed and parameterized forms of some recent RELU's alternatives being the so-called MISH \cite{mish20} and SWISH \cite{swish17} (SWISH) activation functions. 
We recall that the tunable forms of these functions that are given respectively for the 
\olive{Parametric MISH (PMISH)} by \cite{mish20}: 
\begin{IEEEeqnarray}{rCl}
M(x)&=& x \tanh( \softplus_\xi (x) )  \nonumber \\
&=& 
 x \tanh\left( \frac{1}{\xi} \log (1 + \exp{\xi x} ) \right) 
\label{eqmish}
\end{IEEEeqnarray}
and for the 
\brown{Parametric SWISH (PSWISH)} used in \cite{swish17} and also called SiLU (Sigmoid Linear Unit) in \cite{silu20}:
\begin{equation}
S(x)= x \sigma(\xi x) = 
\frac{x}{1+\exp{-\xi x}}
\label{eqswish}
\end{equation}
We will keep the terminologies of MISH and SWISH in the standard cases corresponding to $\xi = 1$ in Eqs. \eqref{eqmish} and \eqref{eqswish}.
We will also consider learning parameter $\xi$ involved in PMISH and PSWISH together with learning standard convolutional weights.

Tables \ref{tab cnnbasic} and \ref{tab cnndeep} respectively provide two shallow (CNN-1-ReSKU and CNN-1-ReLU) and two deep (CNN-2-ReSKU and CNN-2-ReLU) networks to be used in the experimental tests, in addition with their MISH, PMISH, SWISH and PSWISH variants. 
Concerning the number of learnable parameters: the ReSKU based CNNs involve $3 N_2$ additional nonlinear parameters whereas the PMISH/PSWISH based CNNs involve $ N_2$ additional nonlinear parameters, in comparison with the ReLU, MISH and SWISH based CNNs, where $N_2$ is the number of convolution filters used at layer 2.

\begin{table}[!htpb]
\centering
\caption{
Shallow {CNN-1-$X$} frameworks where $X\in$ \{ ReLU, ReSKU, MISH, PMISH, SWISH, PSWISH \} corresponds to the specific activation function used at layer 4. 
FC denotes a \emph{Fully Connected} layer.
We have considered $N_2 = 32$ for CNN-1.
}
\begin{footnotesize}
\begin{tabular}{||@{}c@{}|l@{~}c@{~}c@{~}c@{}||}\hline\hline
Layer    & Content     & $\# N$ of Elements      & Element size     &      \!\!\! \!\!\!  Learnable  \\\hline
     1 &  $\large{\substack{~\\ \textrm{Inputs} \\~\\ \textrm{Images}  \\ ~}}$  &   $N_1 $     &      $M_1^x \times M_1^y \times M_1^c$            & No   \\\hline
     2 &  `Convolve'           &        $N_2$      &                $M_2^x \times M_2^y $                                 & Yes \\\hline
     3  & `Normalize'        &       \multicolumn{3}{c||}{   Standard / Mini-batch        }   \\\hline
     4  &  
     $\large{\substack{
     \\ 
     \textrm{ \red{`ReLU'}~}  \\ \\ 
     \textrm{\blue{`ReSKU'}} \\ \\ 
     \textrm{ \green{`MISH'}~}  \\ \\ 
     \textrm{ \olive{`PMISH'}~}  \\ \\ 
     \textrm{ \magenta{`SWISH'}~}   \\ \\ 
     \textrm{ \brown{`PSWISH'}~} 
     \\~~ }}$  
     &   $ N_2 $        &       $\large{\substack{
     \\ 
     \textrm{\red{-}}  \\ \\ 
     \textrm{\blue{3}} \\ \\ 
     \textrm{\green{-}}  \\ \\ 
     \textrm{\olive{1}}  \\ \\ 
     \textrm{\magenta{-}}  \\ \\ 
     \textrm{\brown{1}}  
     \\~~ }}$   & 
     $\large{\substack{
     \\ 
     \textrm{\red{No}}  \\ \\ 
     \textrm{\blue{Yes}} \\ \\ 
     \textrm{\green{No}}   \\ \\ 
     \textrm{\olive{Yes}} \\ \\ 
     \textrm{\magenta{No}}   \\ \\ 
     \textrm{\brown{Yes}} 
     \\~~ }}$    \\\hline
    5 &  `FC'         &    \multicolumn{3}{c||}{     [Output size $L$ ] }\\\hline
    6 &  `Softmax'    &   \multicolumn{3}{c||}{     Probabilities with respect to $L$ outputs }\\\hline
    7  & `Classify'  &   \multicolumn{3}{c||}{     Cross-entropy (Output: category)   } \\\hline\hline
\end{tabular}
\end{footnotesize}
\label{tab cnnbasic}
\end{table}

\begin{table}[!htpb]
\centering
\caption{
Deep 
{CNN-2-$Y$} frameworks where $Y\in$ \{ ReLU, ReSKU, MISH, PMISH, SWISH, PSWISH \} corresponds to the specific activation function used at layer 4. 
Only one learnable ReSKU layer is used in order to limit computational complexity. FC denotes a \emph{Fully Connected} layer.
We have considered $N_2 = 96$ for CNN-2.
}
\begin{footnotesize}
\begin{tabular}{||@{}r@{~}|l@{}c@{~}c@{~}c@{}||}\hline\hline
Layer    & Content     & $\# N$ of Elements      & Element size     &      \!\!\! \!\!\!  Learnable  \\\hline
     1 &  $\large{\substack{~\\ \textrm{`Inputs'} \\~\\ \textrm{(images)}  \\ ~}}$  &   $N_1 $     &      $M_1^x \times M_1^y \times M_1^c$            & No   \\\hline\hline
     2 &  `Convolve-1'           &        $ N_2 $      &                $3 \times 3 $                                 & Yes \\\hline
     3  & `Normalize-1'        &       \multicolumn{3}{c||}{   Standard / Mini-batch        }   \\\hline
     4  & 
     $\large{\substack{
     \\ 
     \textrm{ \red{`ReLU'}~}  \\ \\ 
     \textrm{\blue{`ReSKU'}} \\ \\ 
     \textrm{ \green{`MISH'}~}  \\ \\ 
     \textrm{ \olive{`PMISH'}~}  \\ \\ 
     \textrm{ \magenta{`SWISH'}~}   \\ \\ 
     \textrm{ \brown{`PSWISH'}~} 
     \\~~ }}$  
     &   $ N_2 $        &       $\large{\substack{
     \\ 
     \textrm{\red{-}}  \\ \\ 
     \textrm{\blue{3}} \\ \\ 
     \textrm{\green{-}}  \\ \\ 
     \textrm{\olive{1}}  \\ \\ 
     \textrm{\magenta{-}}  \\ \\ 
     \textrm{\brown{1}}  
     \\~~ }}$   & 
     $\large{\substack{
     \\ 
     \textrm{\red{No}}  \\ \\ 
     \textrm{\blue{Yes}} \\ \\ 
     \textrm{\green{No}}   \\ \\ 
     \textrm{\olive{Yes}} \\ \\ 
     \textrm{\magenta{No}}   \\ \\ 
     \textrm{\brown{Yes}} 
     \\~~ }}$ 
       \\\hline\hline
     5 &  `Convolve-2'           &        128      &                $5 \times 5 $                                 & Yes \\\hline
     6  & `Normalize-2'        &       \multicolumn{3}{c||}{   Standard / Mini-batch        }   \\\hline
     7  &  `ReLU'  &           &        3   & No    \\\hline\hline
     8 &  `Convolve-3'           &        384      &                $7 \times 7 $                                 & Yes \\\hline
     9  & `Normalize-3'        &       \multicolumn{3}{c||}{   Standard / Mini-batch        }   \\\hline
     10  &  `ReLU'  &           &        3   & No    \\\hline\hline
     11 &  `Convolve-4'           &        192      &                $5 \times 5 $                                 & Yes \\\hline
     12  & `Normalize-4'        &       \multicolumn{3}{c||}{   Standard / Mini-batch        }   \\\hline
     13  &  `ReLU'  &           &        3   & No    \\\hline\hline
     11 &  `Convolve-5'           &        128     &                $3 \times 3 $                                 & Yes \\\hline
     12  & `Normalize-5'        &       \multicolumn{3}{c||}{   Standard / Mini-batch        }   \\\hline
     13  &  `ReLU'  &           &        3   & No    \\\hline\hline
    5 &  `FC-1'         &    \multicolumn{3}{c||}{     [Output size: $4096$ ] }\\\hline
    13  &  `ReLU'  &   $ 4096 $        &        3   & No    \\\hline\hline
    5 &  `FC-2'         &    \multicolumn{3}{c||}{     [Output size: $L$ ] }\\\hline
    6 &  `Softmax'    &   \multicolumn{3}{c||}{     Probability with respect to $L$ outputs }\\\hline
    7  & `Classify'  &   \multicolumn{3}{c||}{     Cross-entropy (Output: category)   } \\\hline\hline
\end{tabular}
\end{footnotesize}
\label{tab cnndeep}
\end{table}


{
\newcolumntype{g}{>{\columncolor[gray]{.95}[.5\tabcolsep]}c}  
\begin{table*}[!ht]
\centering
\caption{
Mean accuracies in percentages over 100 Monte Carlo trials of the handwritten digit recognition issue: impact of the number of epochs, the Convolution Filter Size (CFS) and Number of Convolution Filters (NCF) with respect to 
shallow {CNN-1-$X$} frameworks defined in Table \ref{tab cnnbasic} where $X\in$ \{ ReLU, RePSU, MISH, PMISH, SWISH, PSWISH \}.
}
\begin{tabular}{@{}c@{}c||c@{}c@{}}\hline\hline\hline
\multicolumn{2}{c}{\textbf{Non-learnable activations}} & \multicolumn{2}{c}{\textbf{Learnable activations}} \\\hline\hline
\multicolumn{4}{c}{\textbf{EPOCH = 1}}\\
\begin{tabular}{c}
\begin{turn}{90}  \red{CNN-1-ReLU} \quad \qquad \end{turn}
\end{tabular}
&
\red{
\begin{tabular}{@{}c@{~}|c@{~~}g@{~~}c@{~~}g@{~~}c@{~~}g@{~}}
\backslashbox{NCF}{CFS}
      &    2    &   3    &   4     &   5     &   6    &    7\\\hline     10
         &         72.67         &         72.70         &         72.23         &         72.35         &         72.40         &         72.25     \\     20
         &         77.08         &         77.57         &         77.45         &         77.10         &         77.40         &         77.47     \\     30
         &         78.27         &         78.18         &         78.34         &         78.07         &         78.28         &         77.85     \\     40
         &         77.96         &         77.97         &         77.79         &         77.93         &         77.72         &         77.55     \\     50
         &         77.62         &         77.76         &         77.62         &         77.36         &         77.62         &         77.29     \\     
\end{tabular}}
&
\blue{
\begin{tabular}{@{}c@{~}|c@{~~}g@{~~}c@{~~}g@{~~}c@{~~}g@{~}}
 \backslashbox{NCF}{CFS}
      &    2    &   3    &   4     &   5     &   6    &    7\\\hline     10
         &         78.07         &         76.92         &         78.07         &         77.48         &         76.99         &         77.03     \\     20
         &         84.99         &         85.84         &         83.92         &         85.48         &         84.91         &         84.21     \\     30
         &         86.75         &         87.91         &         88.42         &         88.49         &         88.69         &         85.83     \\     40
         &         89.73         &         88.91         &         87.09         &         88.81         &         88.99         &         88.16     \\     50
         &         89.74         &         87.55         &         89.71         &         89.74         &         90.62         &         \black{{\bf 90.55}}     \\     
\end{tabular}}
&
\begin{tabular}{c}
\begin{turn}{90}  \blue{CNN-1-RePSU} \quad \qquad \end{turn}
\end{tabular}
\\\hline
\begin{tabular}{c}
\begin{turn}{90}  \green{CNN-1-\misho} \quad \qquad \end{turn}
\end{tabular}
&
\green{
\begin{tabular}{@{}c@{~}|c@{~~}g@{~~}c@{~~}g@{~~}c@{~~}g@{~}}
\backslashbox{NCF}{CFS}
      &    2    &   3    &   4     &   5     &   6    &    7\\\hline     10
         &         75.14         &         75.13         &         75.64         &         74.98         &         75.03         &         74.83     \\     20
         &         81.69         &         81.83         &         81.85         &         81.51         &         81.80         &         81.44     \\     30
         &         84.24         &         84.19         &         84.13         &         84.12         &         83.93         &         84.07     \\     40
         &         84.88         &         85.02         &         84.62         &         84.83         &         85.00         &         84.75     \\     50
         &         84.89         &         84.96         &         84.80         &         84.69         &         85.01         &         84.93     \\     
\end{tabular}}
&
\olive{
\begin{tabular}{@{}c@{~}|c@{~~}g@{~~}c@{~~}g@{~~}c@{~~}g@{~}}
 \backslashbox{NCF}{CFS}
      &    2    &   3    &   4     &   5     &   6    &    7\\\hline     10
         &         66.38         &         66.60         &         66.71         &         66.98         &         66.68         &         66.77     \\     20
         &         68.68         &         68.42         &         68.84         &         68.29         &         68.46         &         68.59     \\     30
         &         68.55         &         69.00         &         68.72         &         68.47         &         68.46         &         68.52     \\     40
         &         68.27         &         68.02         &         67.96         &         68.02         &         68.00         &         68.07     \\     50
         &         67.28         &         67.57         &         67.22         &         66.89         &         67.11         &         67.55     \\     
\end{tabular}}
&
\begin{tabular}{c}
\begin{turn}{90}  \olive{CNN-1-\mish} \quad \qquad \end{turn}
\end{tabular}
\\\hline
\begin{tabular}{c}
\begin{turn}{90}  \magenta{CNN-1-\swisho} \quad \qquad \end{turn}
\end{tabular}
&
\magenta{
\begin{tabular}{@{}c@{~}|c@{~~}g@{~~}c@{~~}g@{~~}c@{~~}g@{~}}
\backslashbox{NCF}{CFS}
      &    2    &   3    &   4     &   5     &   6    &    7\\\hline     10
         &         75.40         &         75.19         &         75.11         &         74.98         &         75.16         &         75.27     \\     20
         &         82.17         &         82.03         &         81.82         &         81.89         &         82.23         &         81.78     \\     30
         &         84.70         &         84.46         &         84.33         &         84.29         &         84.57         &         84.63     \\     40
         &         85.51         &         85.47         &         85.48         &         85.41         &         85.42         &         85.57     \\     50
         &         85.82         &         85.76         &         85.81         &         85.46         &         85.65         &         85.89     \\     
\end{tabular}}
&
\brown{
\begin{tabular}{@{}c@{~}|c@{~~}g@{~~}c@{~~}g@{~~}c@{~~}g@{~}}
 \backslashbox{NCF}{CFS}
      &    2    &   3    &   4     &   5     &   6    &    7\\\hline     10
         &         70.23         &         70.37         &         70.14         &         70.16         &         70.49         &         70.79     \\     20
         &         75.81         &         75.79         &         76.19         &         75.71         &         75.75         &         76.08     \\     30
         &         78.19         &         78.06         &         77.80         &         78.20         &         77.86         &         77.98     \\     40
         &         78.55         &         78.76         &         78.63         &         78.87         &         78.59         &         78.61     \\     50
         &         79.27         &         79.29         &         78.81         &         79.41         &         79.12         &         79.24     \\     
\end{tabular}}
&
\begin{tabular}{c}
\begin{turn}{90}  \brown{CNN-1-\swish} \quad \qquad \end{turn}
\end{tabular}
\\\hline
\hline
\multicolumn{4}{c}{\textbf{EPOCH = 2}}\\
\begin{tabular}{c}
\begin{turn}{90}  \red{CNN-1-ReLU} \quad \qquad \end{turn}
\end{tabular}
&
\red{
\begin{tabular}{@{}c@{~}|c@{~~}g@{~~}c@{~~}g@{~~}c@{~~}g@{~}}
\backslashbox{NCF}{CFS}
      &    2    &   3    &   4     &   5     &   6    &    7\\\hline     10
         &         84.37         &         84.07         &         84.11         &         84.34         &         84.48         &         84.58     \\     20
         &         89.10         &         89.09         &         89.23         &         88.94         &         89.15         &         89.58     \\     30
         &         90.56         &         90.46         &         90.79         &         90.47         &         90.65         &         90.54     \\     40
         &         90.93         &         90.99         &         90.86         &         91.22         &         90.95         &         91.11     \\     50
         &         91.12         &         91.18         &         91.15         &         91.15         &         91.07         &         91.03     \\        
\end{tabular}}
&
\blue{
\begin{tabular}{@{}c@{~}|c@{~~}g@{~~}c@{~~}g@{~~}c@{~~}g@{~}}
\backslashbox{NCF}{CFS}
      &    2    &   3    &   4     &   5     &   6    &    7\\\hline     10
         &         88.15         &         88.11         &         88.12         &         88.91         &         88.03         &         88.79     \\     20
         &         93.35         &         91.42         &         89.45         &         91.29         &         94.11         &         90.40     \\     30
         &         94.73         &         93.82         &         90.96         &         90.94         &         93.84         &         91.95     \\     40
         &         94.49         &         88.68         &         91.55         &         88.59         &         92.46         &         94.54     \\     50
         &         92.87         &         91.81         &         90.84         &         91.83         &         94.49         &         \black{{\bf 95.64}}     \\     
\end{tabular}}
&
\begin{tabular}{c}
\begin{turn}{90}  \blue{CNN-1-RePSU} \quad \qquad \end{turn}
\end{tabular}
\\\hline
\begin{tabular}{c}
\begin{turn}{90}  \green{CNN-1-\misho} \quad \qquad \end{turn}
\end{tabular}
&
\green{
\begin{tabular}{@{}c@{~}|c@{~~}g@{~~}c@{~~}g@{~~}c@{~~}g@{~}}
\backslashbox{NCF}{CFS}
      &    2    &   3    &   4     &   5     &   6    &    7\\\hline     10
         &         86.74         &         86.54         &         86.76         &         86.28         &         86.51         &         86.43     \\     20
         &         91.96         &         92.11         &         91.99         &         92.20         &         92.00         &         91.95     \\     30
         &         93.44         &         93.43         &         93.62         &         93.66         &         93.51         &         93.51     \\     40
         &         94.14         &         94.07         &         94.07         &         94.15         &         94.12         &         94.10     \\     50
         &         94.09         &         94.30         &         94.44         &         94.34         &         94.40         &         94.29     \\      
\end{tabular}}
&
\olive{
\begin{tabular}{@{}c@{~}|c@{~~}g@{~~}c@{~~}g@{~~}c@{~~}g@{~}}
\backslashbox{NCF}{CFS}
      &    2    &   3    &   4     &   5     &   6    &    7\\\hline     10
         &         73.67         &         74.18         &         73.47         &         74.28         &         74.52         &         74.37     \\     20
         &         76.85         &         77.06         &         77.37         &         76.92         &         77.03         &         77.07     \\     30
         &         77.90         &         77.28         &         77.61         &         77.40         &         77.53         &         77.64     \\     40
         &         77.70         &         77.61         &         77.25         &         77.59         &         77.04         &         77.37     \\     50
         &         76.80         &         76.63         &         77.11         &         76.95         &         76.59         &         76.82     \\     
\end{tabular}}
&
\begin{tabular}{c}
\begin{turn}{90}  \olive{CNN-1-\mish} \quad \qquad \end{turn}
\end{tabular}
\\\hline
\begin{tabular}{c}
\begin{turn}{90}  \magenta{CNN-1-\swisho} \quad \qquad \end{turn}
\end{tabular}
&
\magenta{
\begin{tabular}{@{}c@{~}|c@{~~}g@{~~}c@{~~}g@{~~}c@{~~}g@{~}}
\backslashbox{NCF}{CFS}
      &    2    &   3    &   4     &   5     &   6    &    7\\\hline     10
         &         86.30         &         86.60         &         86.47         &         86.34         &         86.58         &         86.37     \\     20
         &         92.10         &         91.99         &         92.09         &         91.87         &         91.68         &         91.95     \\     30
         &         93.64         &         93.63         &         93.63         &         93.58         &         93.64         &         93.67     \\     40
         &         94.22         &         94.28         &         94.26         &         94.39         &         94.18         &         94.40     \\     50
         &         94.54         &         94.58         &         94.54         &         94.52         &         94.68         &         94.61     \\       
\end{tabular}}
&
\brown{
\begin{tabular}{@{}c@{~}|c@{~~}g@{~~}c@{~~}g@{~~}c@{~~}g@{~}}
\backslashbox{NCF}{CFS}
      &    2    &   3    &   4     &   5     &   6    &    7\\\hline     10
         &         80.30         &         80.39         &         80.78         &         80.54         &         80.24         &         80.69     \\     20
         &         86.57         &         86.28         &         86.48         &         86.49         &         86.84         &         86.63     \\     30
         &         88.86         &         88.87         &         88.67         &         88.72         &         88.73         &         88.67     \\     40
         &         89.58         &         89.76         &         89.65         &         89.38         &         89.45         &         89.72     \\     50
         &         90.13         &         89.96         &         89.87         &         89.79         &         89.69         &         89.79     \\     
\end{tabular}}
&
\begin{tabular}{c}
\begin{turn}{90}  \brown{CNN-1-\swish} \quad \qquad \end{turn}
\end{tabular}
\\\hline
\hline\hline
\end{tabular}
\label{tab montecarloRePSU}
\end{table*}
}

\newcolumntype{g}{>{\columncolor[gray]{.95}[.5\tabcolsep]}c}  
\begin{table*}[!ht]
\centering
\caption{
Mean validation loss and mean validation accuracy every ten epochs for the GFBF class identification issue with respect to RePSU and ReLU based deep CNN presented in Table \ref{tab cnndeep}.
}
\begin{tabular}{||l@{~}||c@{~~~~}g@{~~~~}c@{~~~~}g@{~~~~}c@{~~~~}g@{~~~~}c@{~~~~}g@{~~~~}||@{~~~~}c@{~~}||}\hline\hline
Max Epochs  & 5 & 10 & 15 & 20 & 25 & 50 & 75 &  100  &   \\\hline\hline
 & \multicolumn{8}{c}{} &   Time elapsed (training) \\
 & \multicolumn{8}{c}{Validation accuracy} &    (hh:mm:ss)  \\\hline\hline
\red{CNN-2-ReLU} &
 53.75 & 59.90  &   60.94 &  64.58  &  63.75 & 58.13  & 58.13  & 58.96
 &
 06:17:10
 \\
\green{CNN-2-\misho} &
54.06 &  58.33  &  60.42   &   62.92  &  69.38   &   71.25   &    66.15    &         66.25 &
06:35:30 \\
\olive{CNN-2-\mish} &
39.58 & 48.12  &  49.38   &   60.31   &    65.31  &   61.88   &   71.04   &   68.54   &
07:18:37
\\
\magenta{CNN-2-\swisho} &
44.17 &  52.40  &   60.31   &  61.35  &  62.60   &   62.81   &   64.17   &    60.73   &
06:25:57
\\
\brown{CNN-2-\swish} &
46.25 &  56.98  &   62.81   &   63.75  &   67.81  &  60.42   &   69.58   &   69.58  &   
06:30:53
\\\hline
\multicolumn{10}{||c||}{RePSU and special cases: RePSKU ($\alpha=0$) and RePSHU ($\alpha=1$)}
\\\hline
\blue{CNN-2-RePSU} &
48.54  & 62.19  &   58.96   &   69.79  &   71.77   &   73.54  &  73.65   &   74.27   &   
08:59:29
 \\
\blue{CNN-2-RePSKU} &
50.10  & 52.60  &   67.08   &   68.02  &   74.69   &   70.83  &  71.77   &   72.40   &   
07:25:43
\\
\blue{CNN-2-RePSHU} &
45.31  & 60.94  &   59.38   &   62.50  &   71.88   &   72.29  &  70.83   &   71.56   &   
07:54:28
 \\
\hline\hline
\end{tabular}
\label{tab gfbfcnnperfo}
\end{table*}

\subsection{
Monte Carlo validation over a handwritten digit recognition problem on shallow CNN 
}\label{sseclearn1}

We consider a standard handwritten digit recognition problem \cite{tensorflow16}, when training and testing concern the shallow CNN described by Table \ref{tab cnnbasic}. The issue addressed in this section is the achievable learning rate when the number of training epoch is fixed to either 1 or 2.
A Monte Carlo simulation framework is proposed to avoid a biased comparison that can be due to sensitivity in random number generation.

100 Monte Carlo iterations have been used for any recognition task associated with the following experimental setup:
\begin{itemize}
\item
Split the handwritten digit database in training and testing sets;
\item
Specify a number of epochs and perform iteratively, the following Monte Carlo experiments:
\begin{itemize}
\item
initialize RePSU, PMISH, PSWISH parameters from positive random numbers, 
\item
perform training\footnote{In RePSU case, the corresponding cross-entropy depends on RePSU form and RePSU parameters: the latter, except $\beta$ fixed to 1 because it is a power term, are updated thanks to the behavior of the cross-entropy and the RePSU derivatives by using the back-propagation algorithm with respect to a gradient descent method, similarly as when updating convolution weights.} with respect to the number of epochs, then testing
\item
save testing score and reiterate;
\end{itemize}
\item
Compute average performance over the correspond 100 Monte Carlo trials.
\end{itemize}
Experimental results are given in Table \ref{tab montecarloRePSU} depending on the number of epochs and the sizing\footnote{The number of RePSU functions used is $N_2$ (number of convolution filters used downstream): one RePSU function per convolution filter.} of layer 2 (numbers and sizes of convolution filters, which determine the number of additional parameters used).
It appears that the RePSU CNN is systematically more performant than the ReLU, MISH, PMISH, SWISH and PSWISH CNNs in terms of the speed in learning a good classifier with respect to the training database.


{
\linespread{0.6}



\begin{figure*}[!t]
\centering
 {Increasing temporal interactions (inter-class dependencies)}
 \\
{%
\begin{tabular}{@{}c@{}c@{}c@{}c@{}c@{}}
{{
\begin{tabular}{@{}c@{}}
\begin{tabular}{@{}c@{}c@{}}
{} 
&
\begin{tabular}{@{}c@{}}
Class 1 \\
$\left\Vert \genfbf_{{\frak H}_{1-\bullet}} \right\Vert $\\
\includegraphics[width=2.40cm]{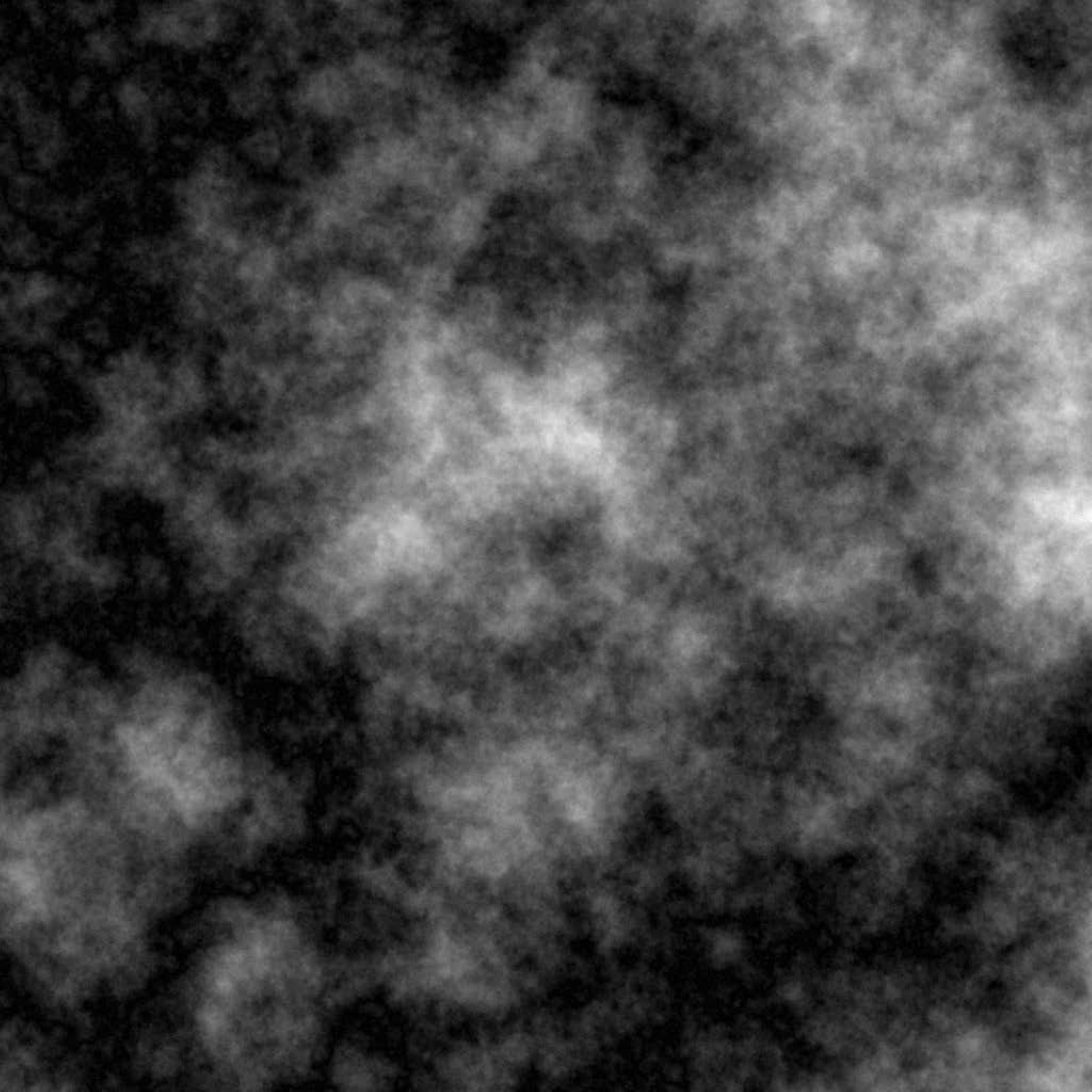}\\
\includegraphics[width=2.40cm]{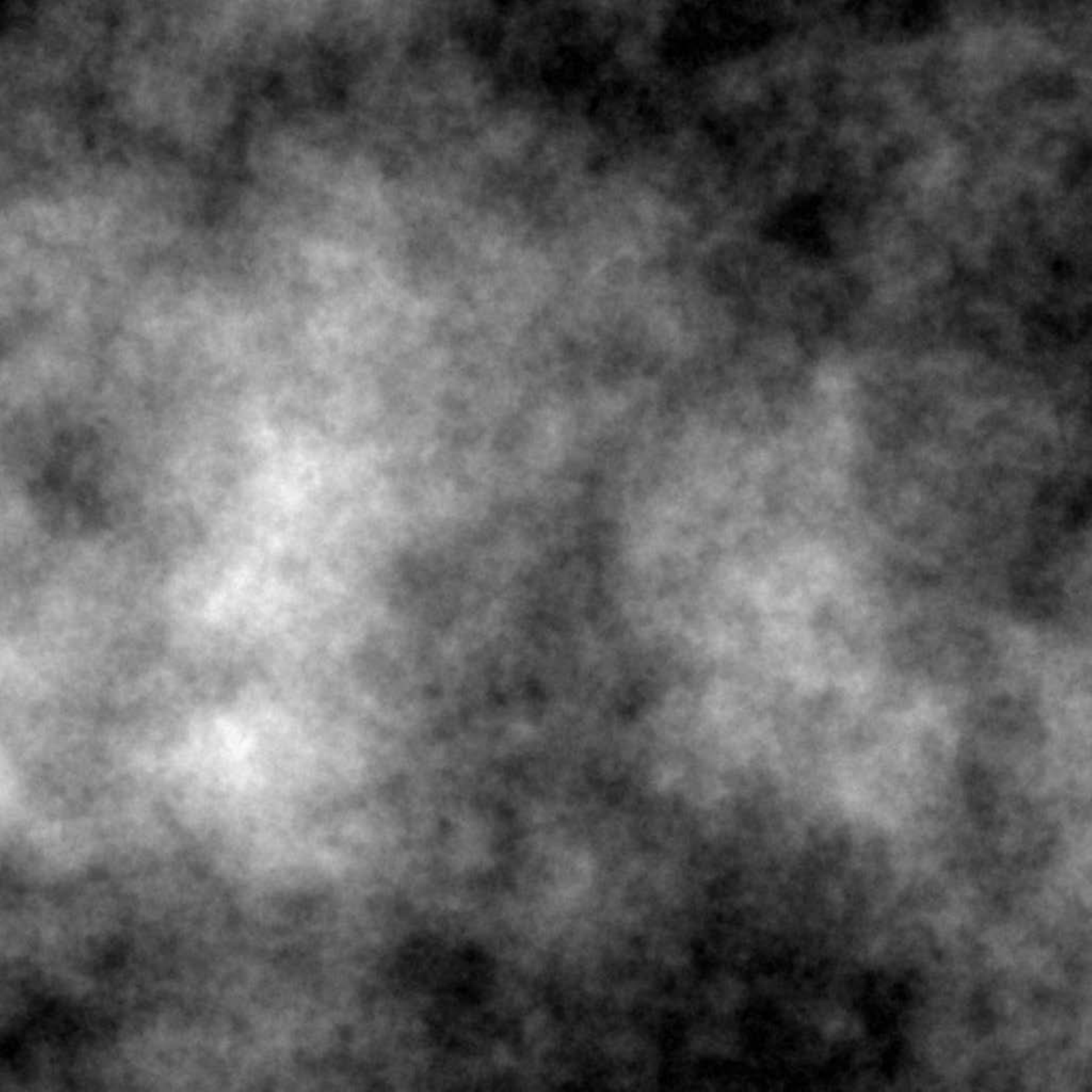}\\
\includegraphics[width=2.40cm]{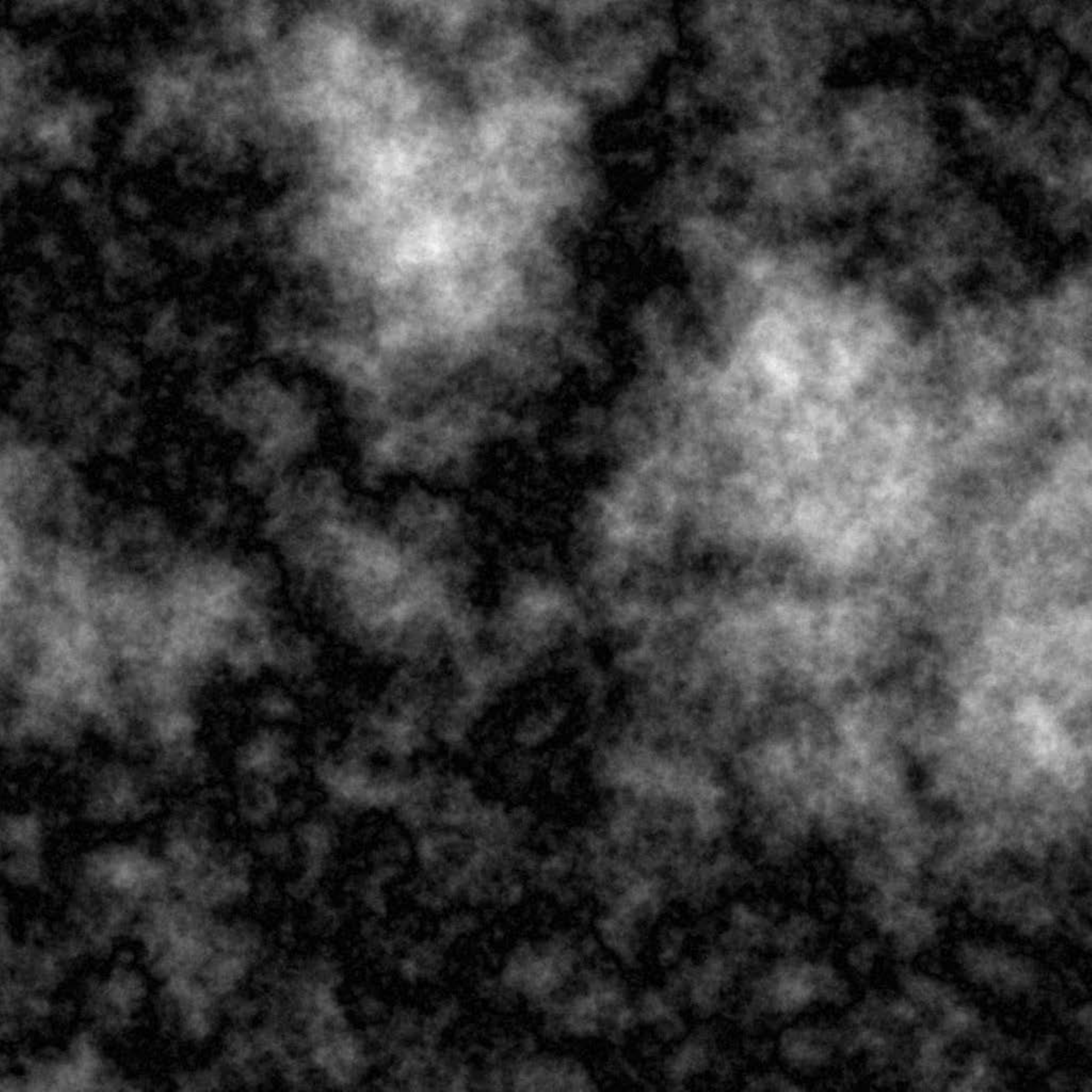}\\
\includegraphics[width=2.40cm]{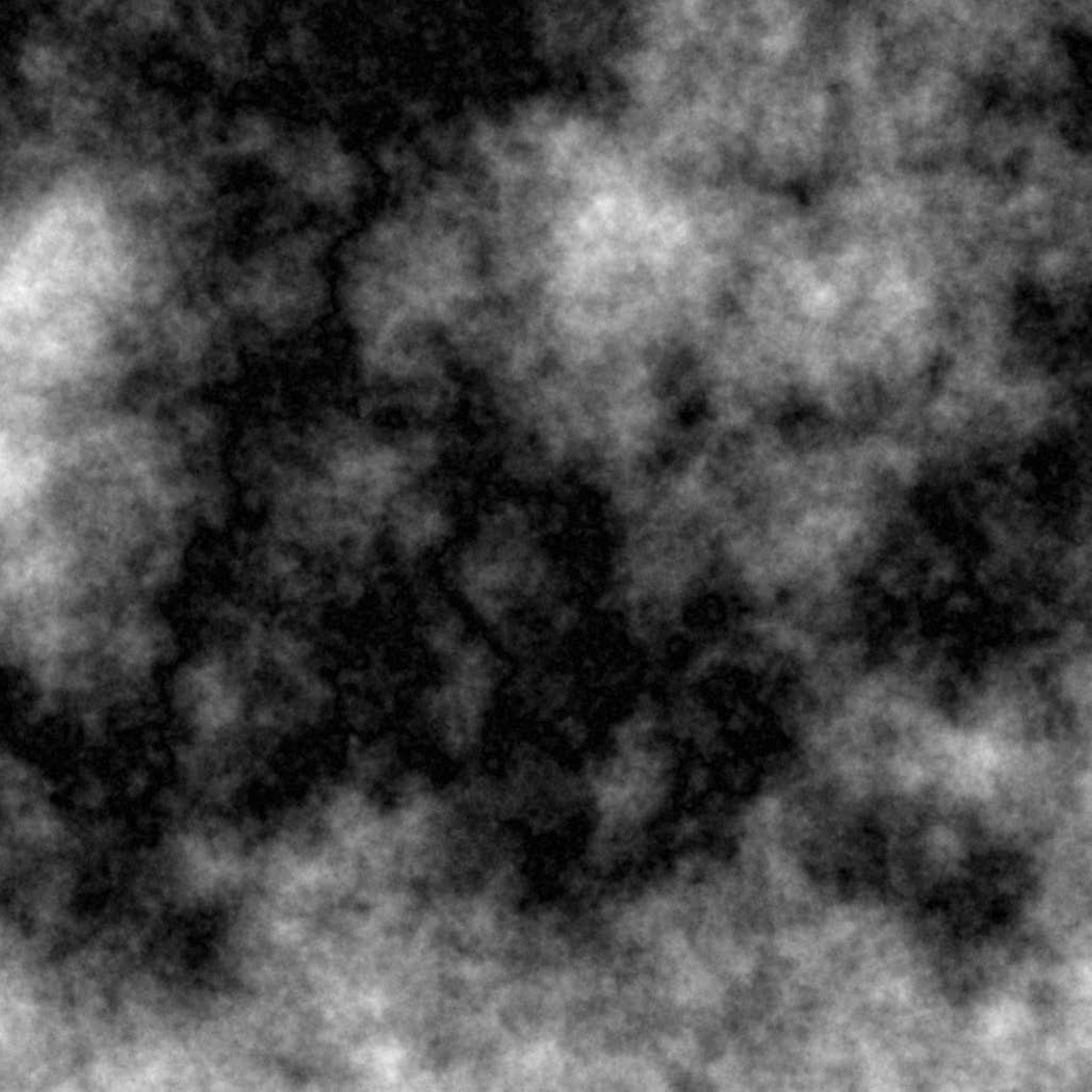}\\
\includegraphics[width=2.40cm]{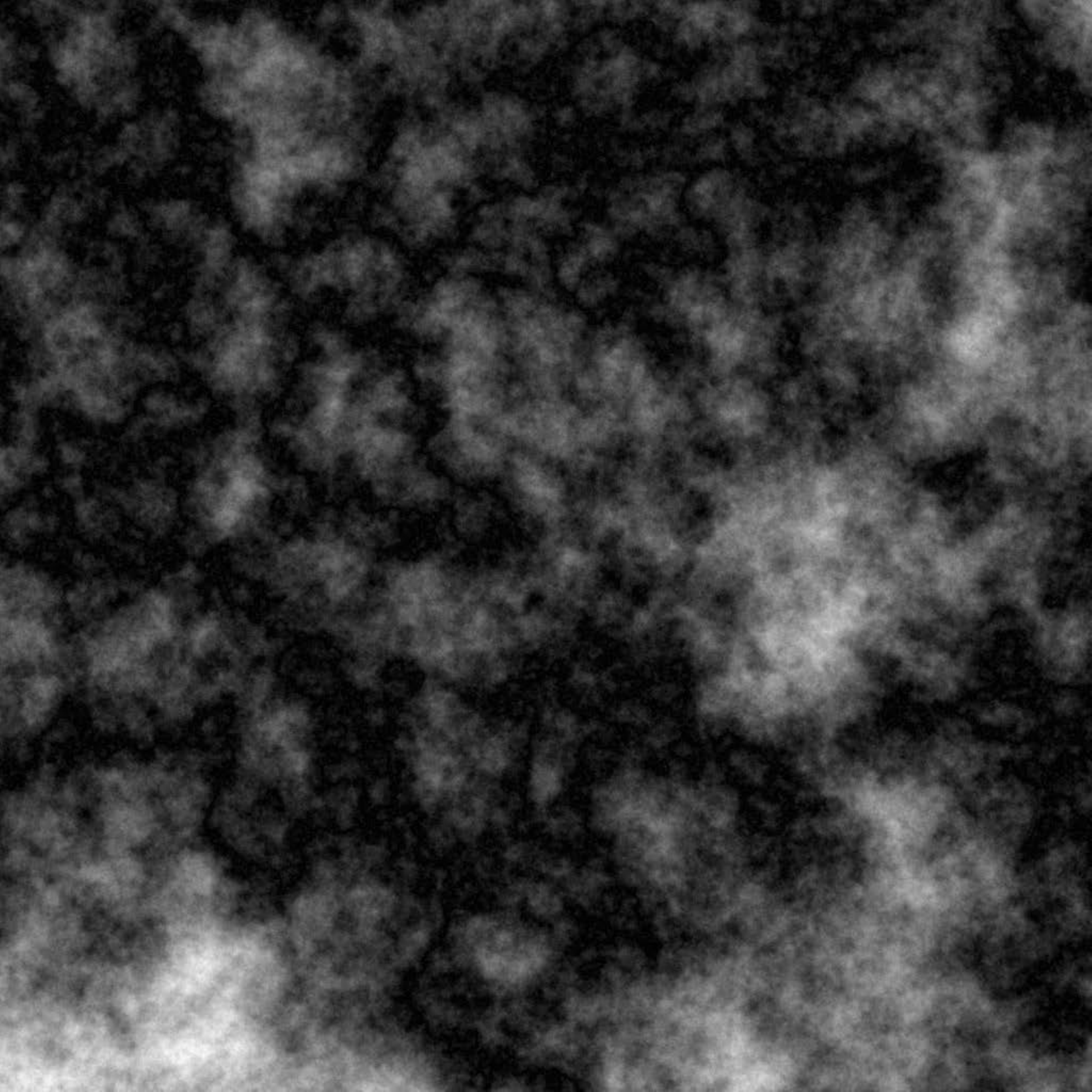}\\
\includegraphics[width=2.40cm]{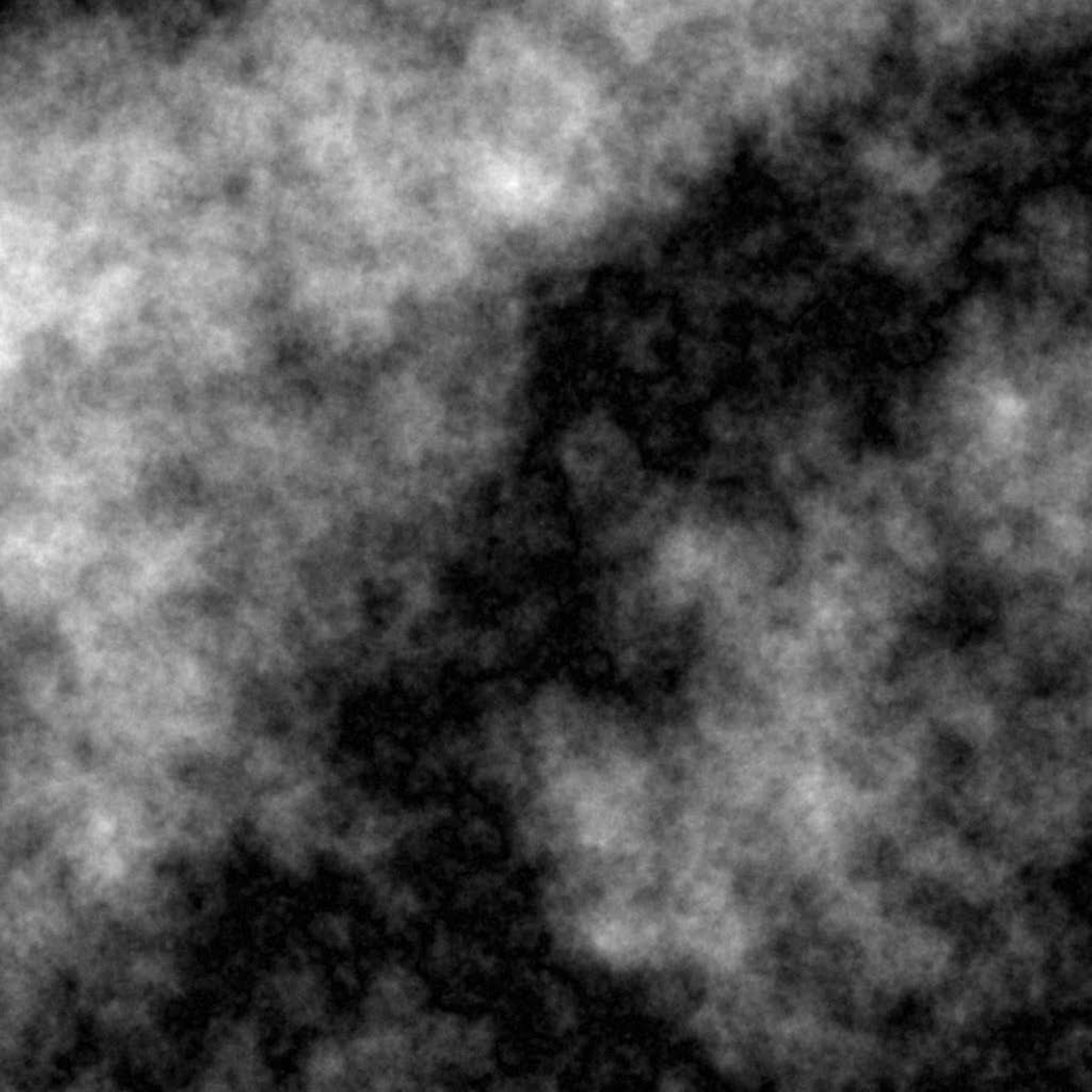}\\
\includegraphics[width=2.40cm]{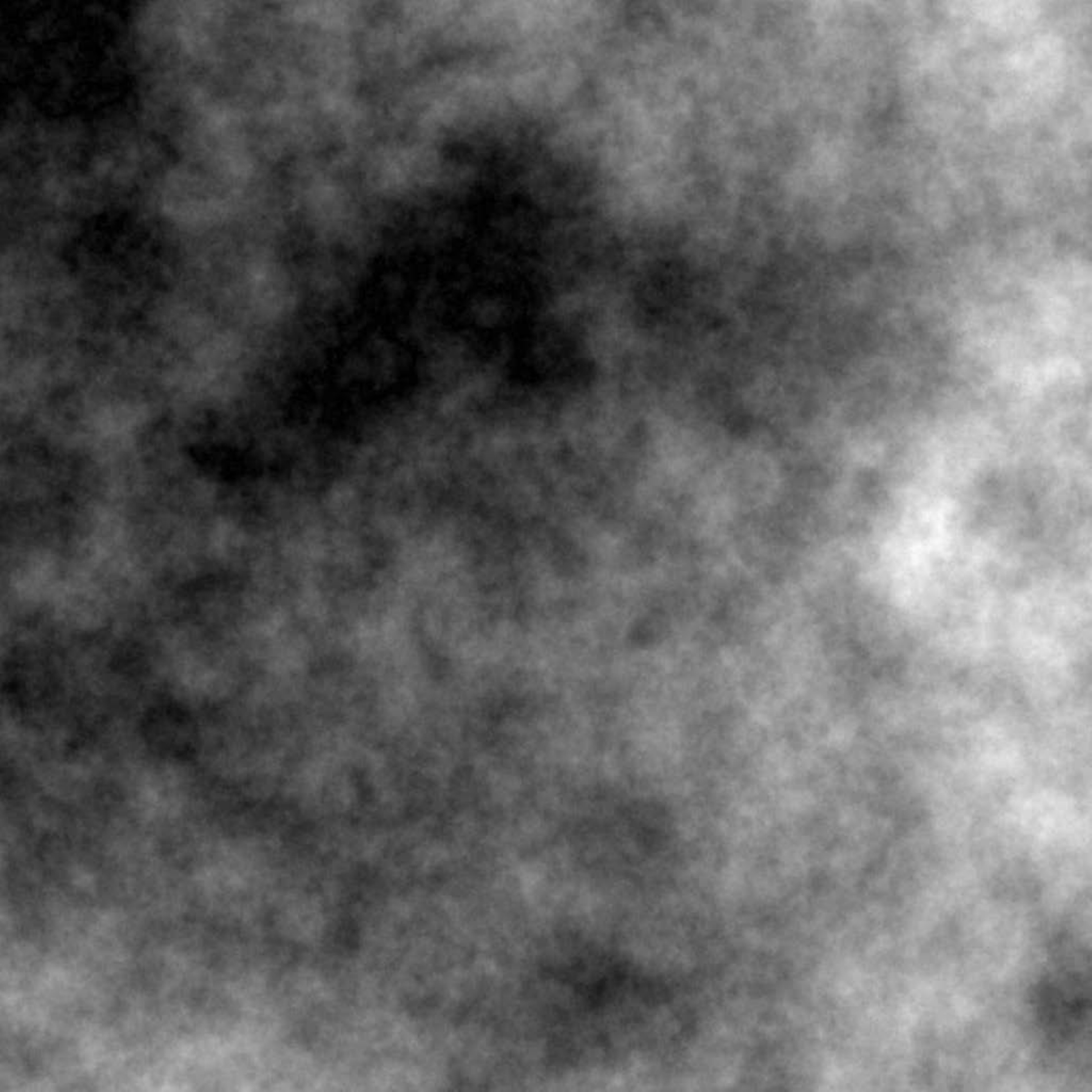}\\
\includegraphics[width=2.40cm]{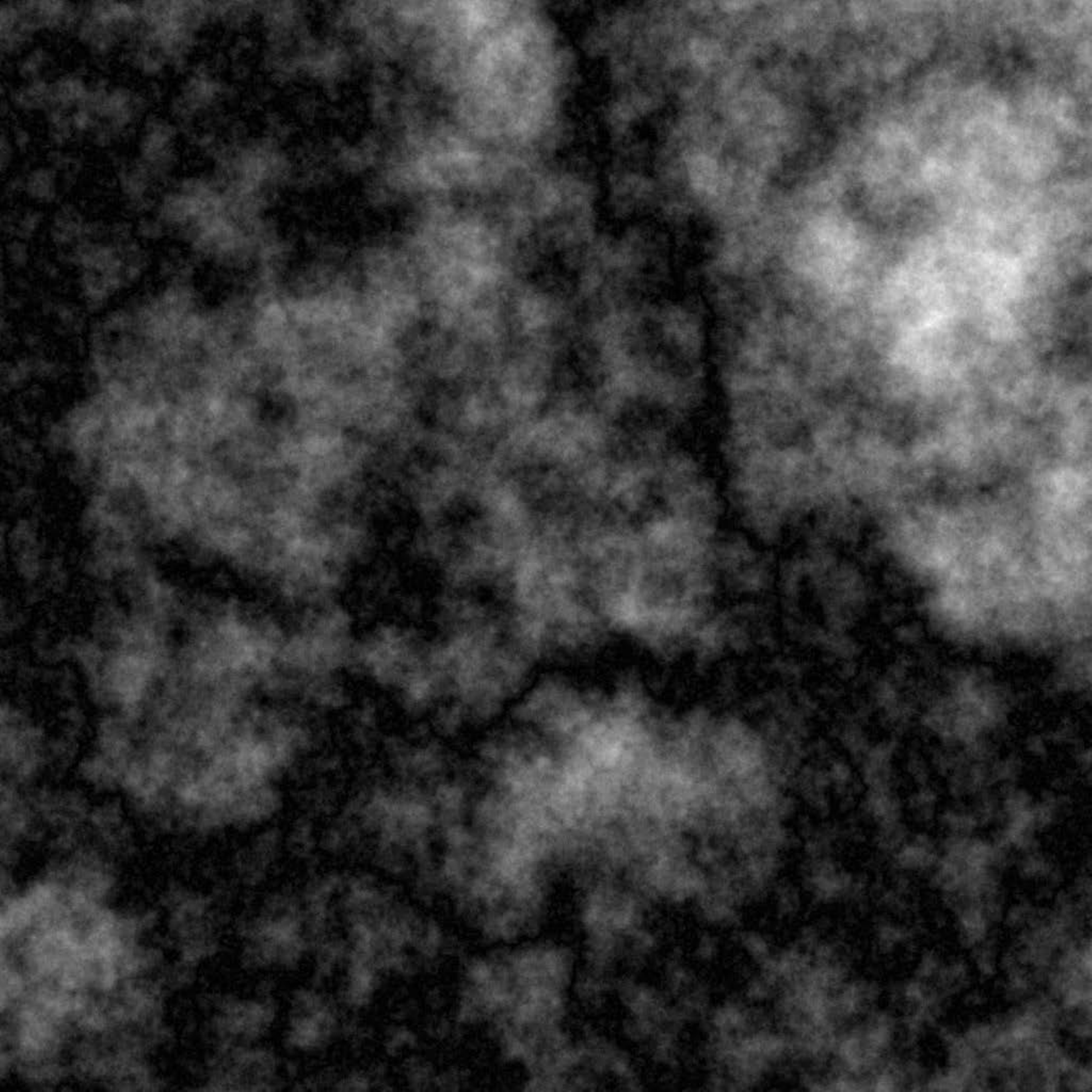}\\
\end{tabular}
\end{tabular}
\end{tabular}
}
}
&
{{
${\!\stackrel{\star}{\Rightarrow}} \:$
\begin{tabular}{@{}c@{}}
\begin{tabular}{@{}c@{}c@{}}
{} 
&
\begin{tabular}{@{}c@{}}
Class 2 \\
$\left\Vert \genfbf_{{\frak H}_{2-\bullet}} \right\Vert $\\
\includegraphics[width=2.40cm]{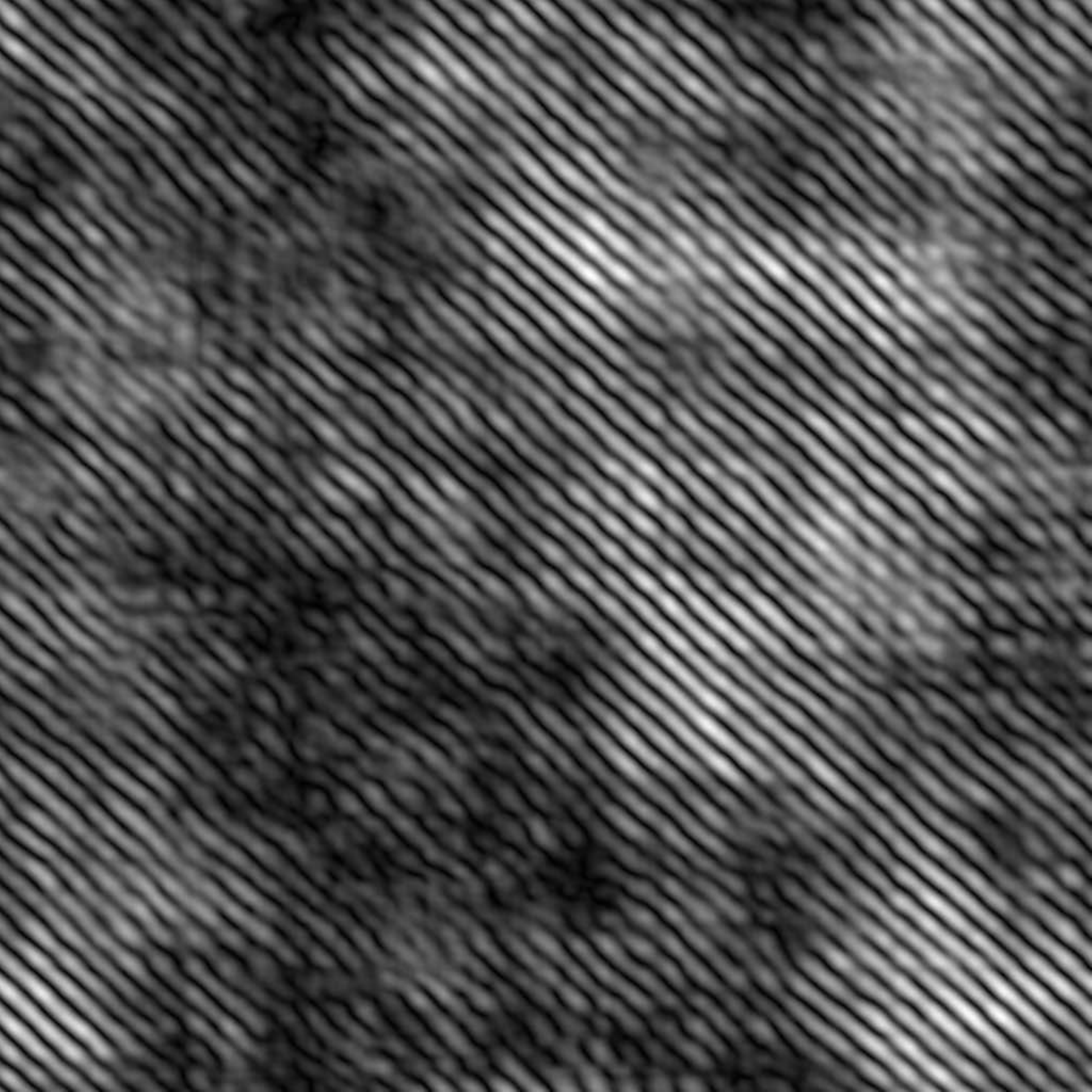}\\
\includegraphics[width=2.40cm]{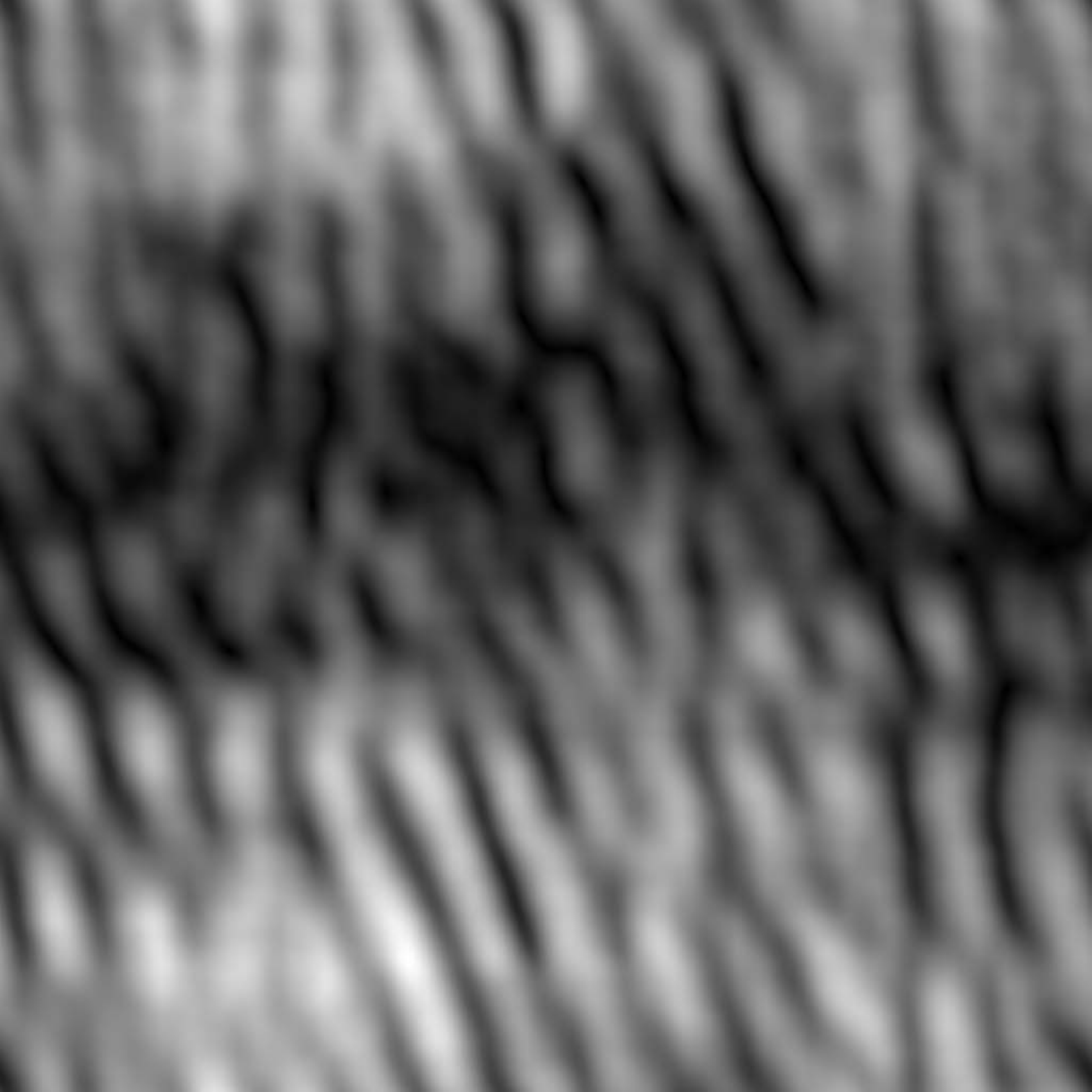}\\
\includegraphics[width=2.40cm]{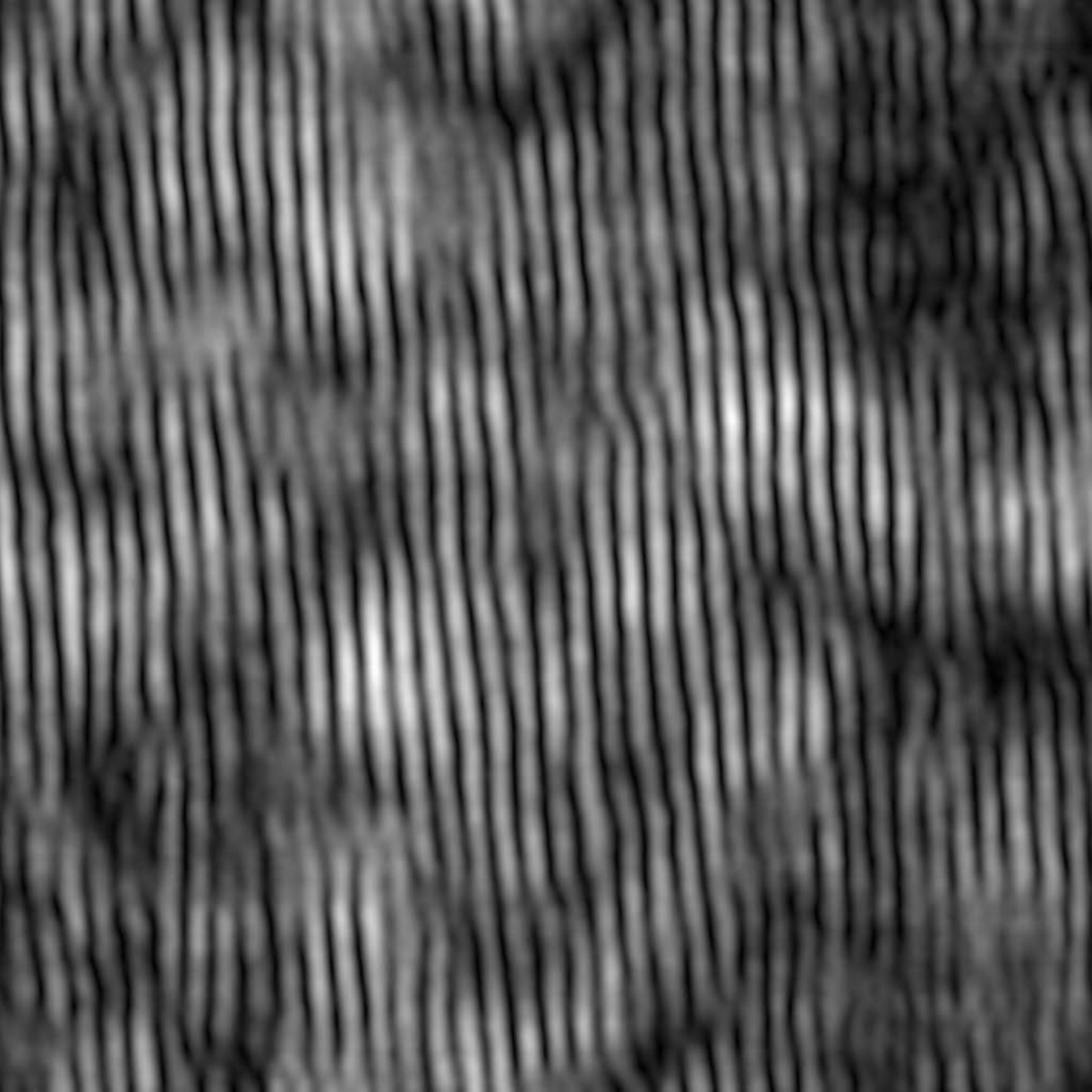}\\
\includegraphics[width=2.40cm]{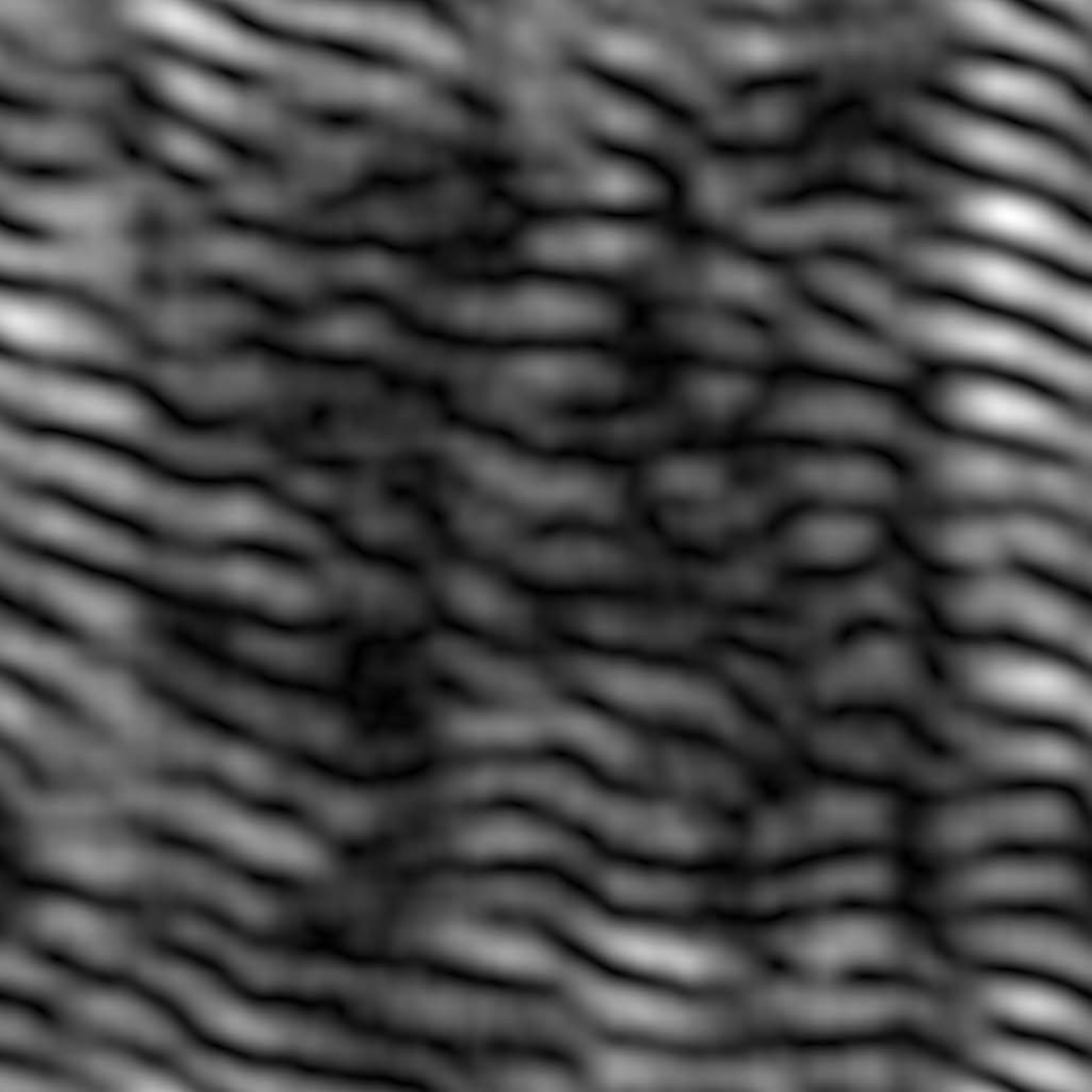}\\
\includegraphics[width=2.40cm]{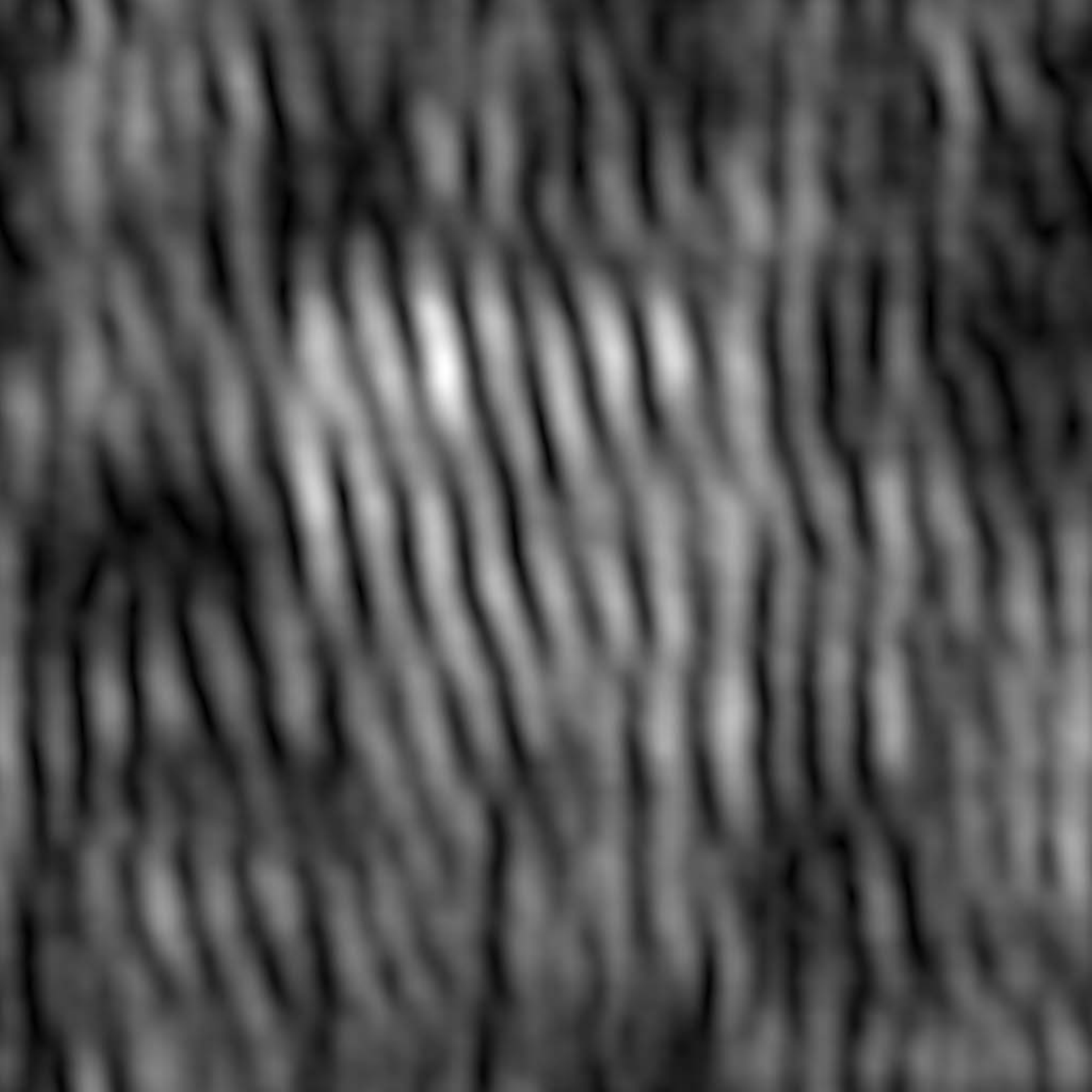}\\
\includegraphics[width=2.40cm]{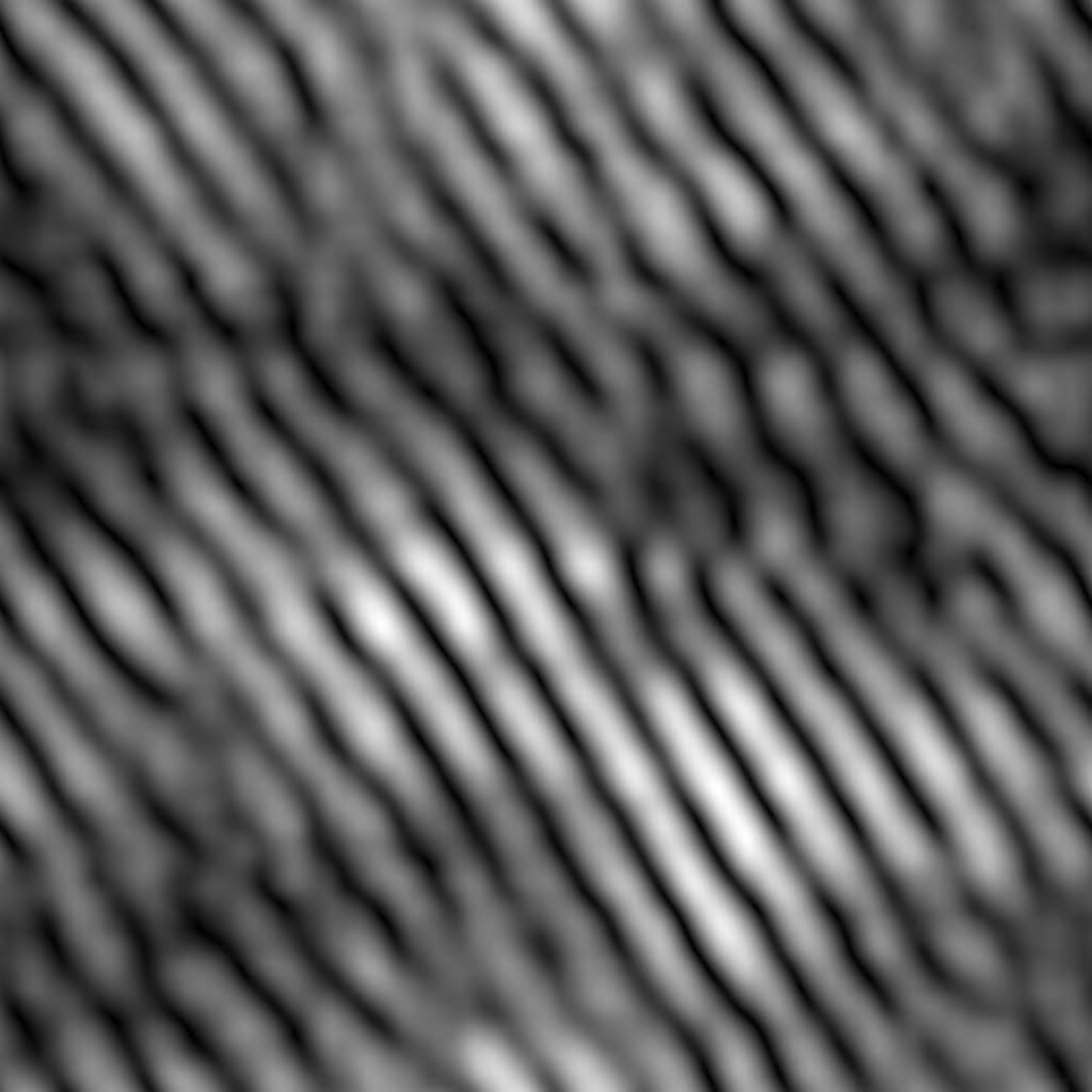}\\
\includegraphics[width=2.40cm]{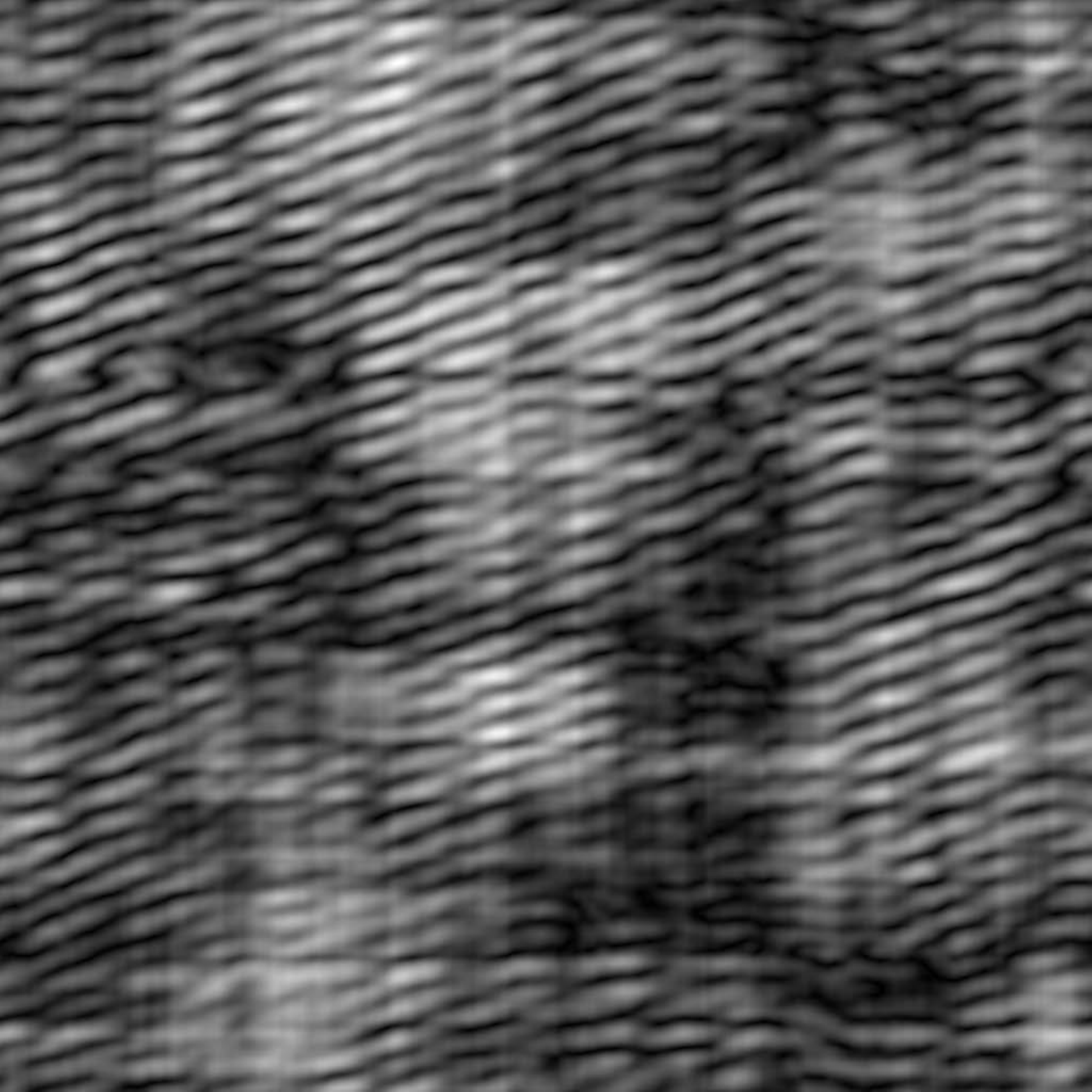}\\
\includegraphics[width=2.40cm]{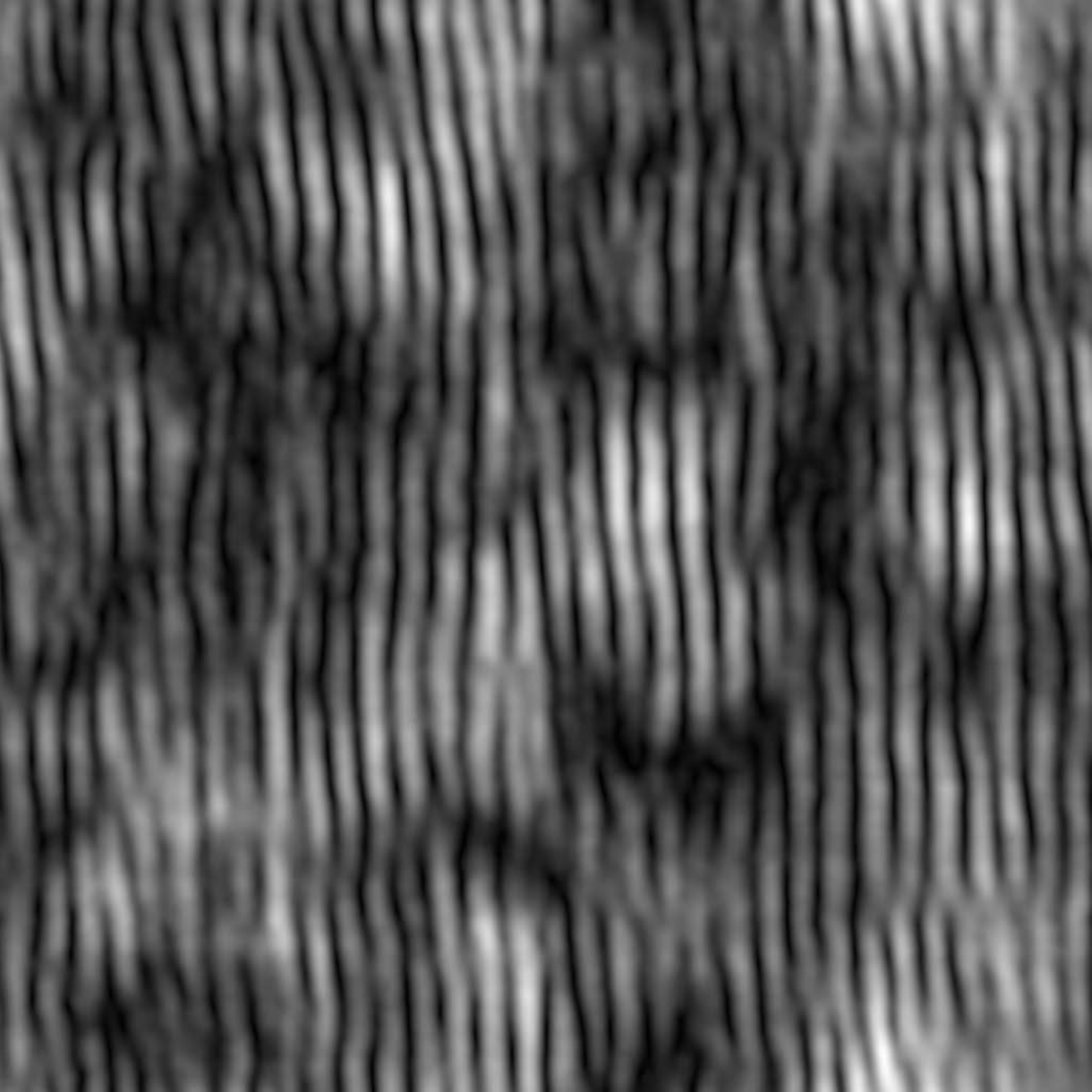}\\
\end{tabular}
\end{tabular}
\end{tabular}
}
}
&
{{
${\!\stackrel{\star}{\Rightarrow}} \:$
\begin{tabular}{@{}c@{}}
\begin{tabular}{@{}c@{}c@{}}
{} 
&
\begin{tabular}{@{}c@{}}
Class 3 \\
$\left\Vert \genfbf_{{\frak H}_{3-\bullet}} \right\Vert $\\
\includegraphics[width=2.40cm]{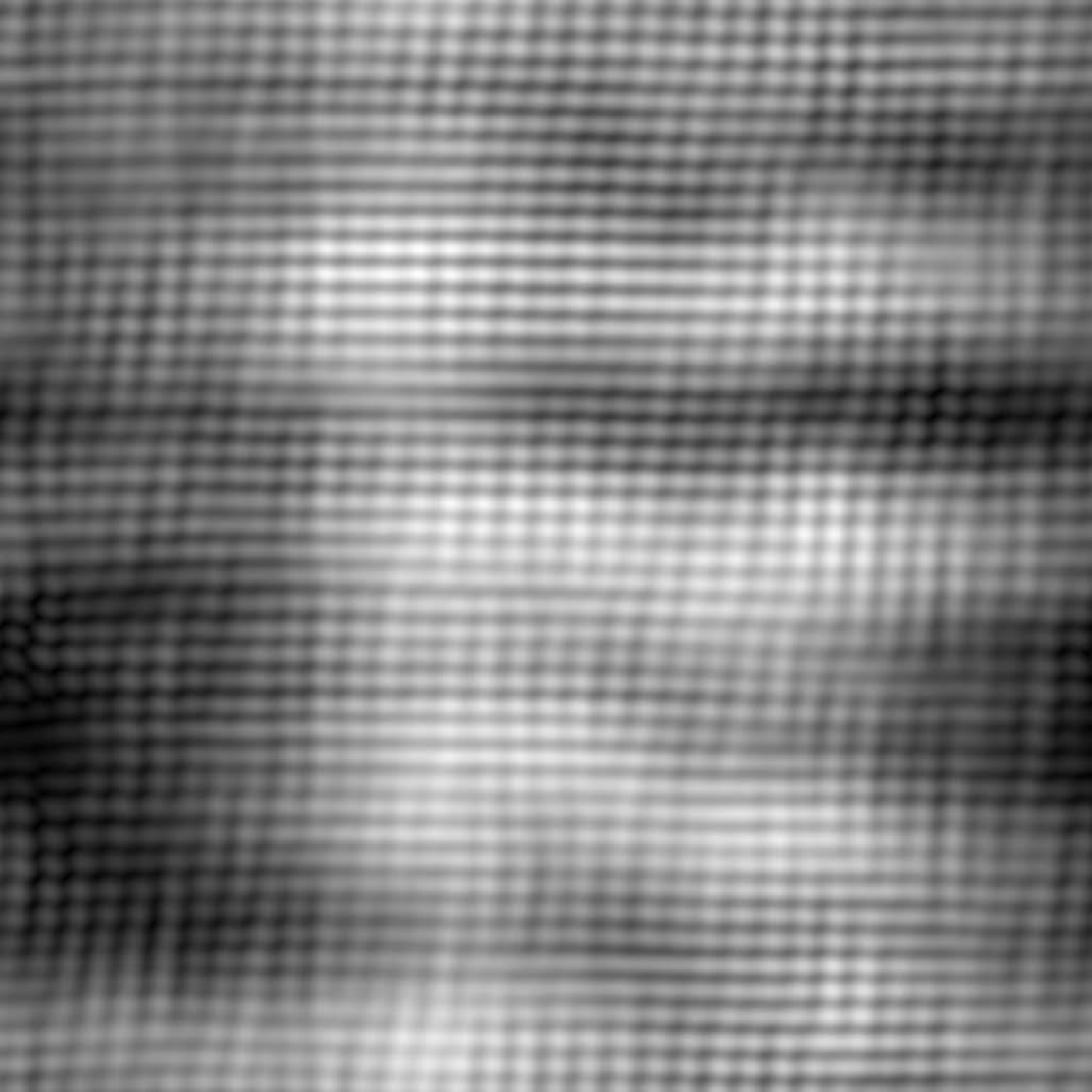}\\
\includegraphics[width=2.40cm]{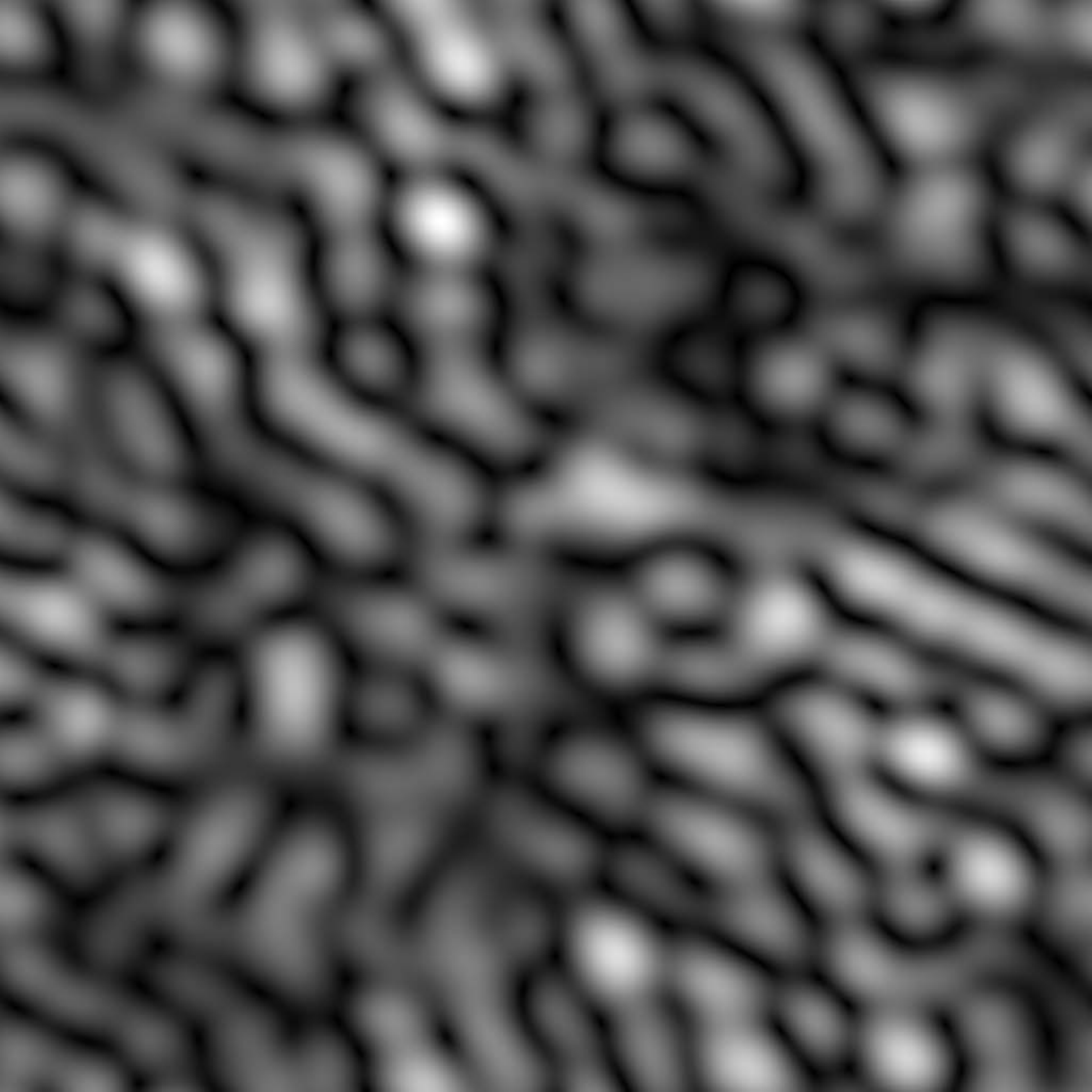}\\
\includegraphics[width=2.40cm]{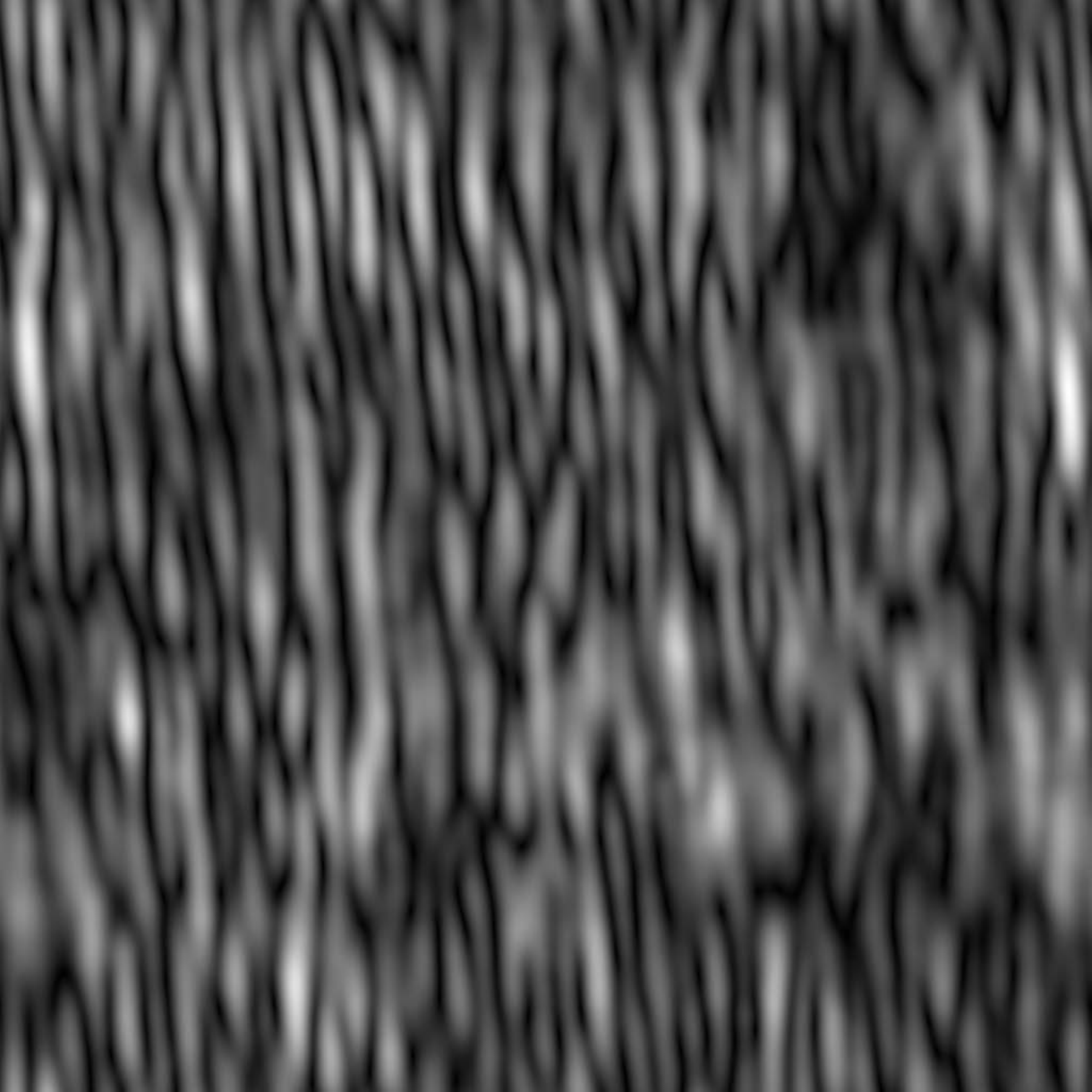}\\
\includegraphics[width=2.40cm]{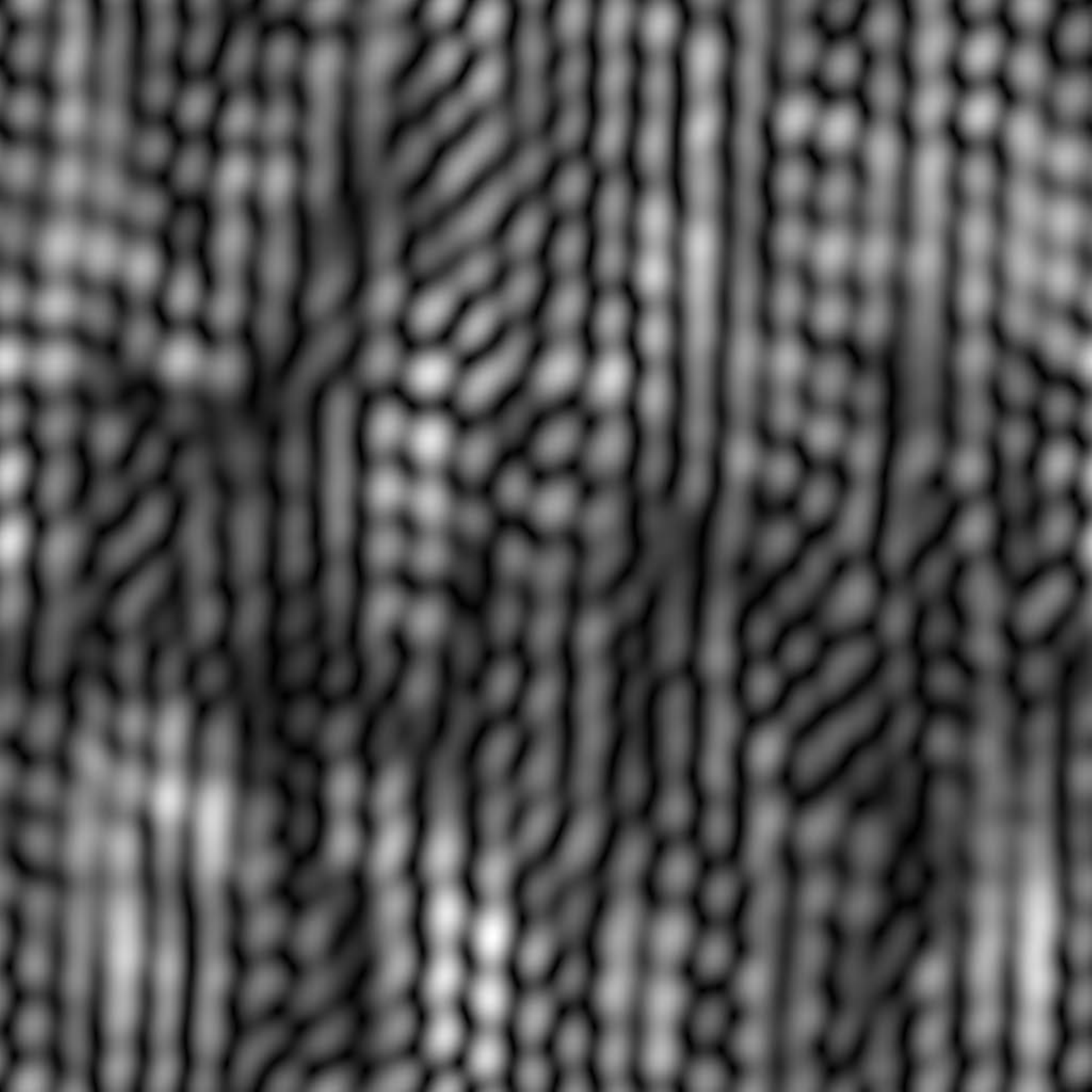}\\
\includegraphics[width=2.40cm]{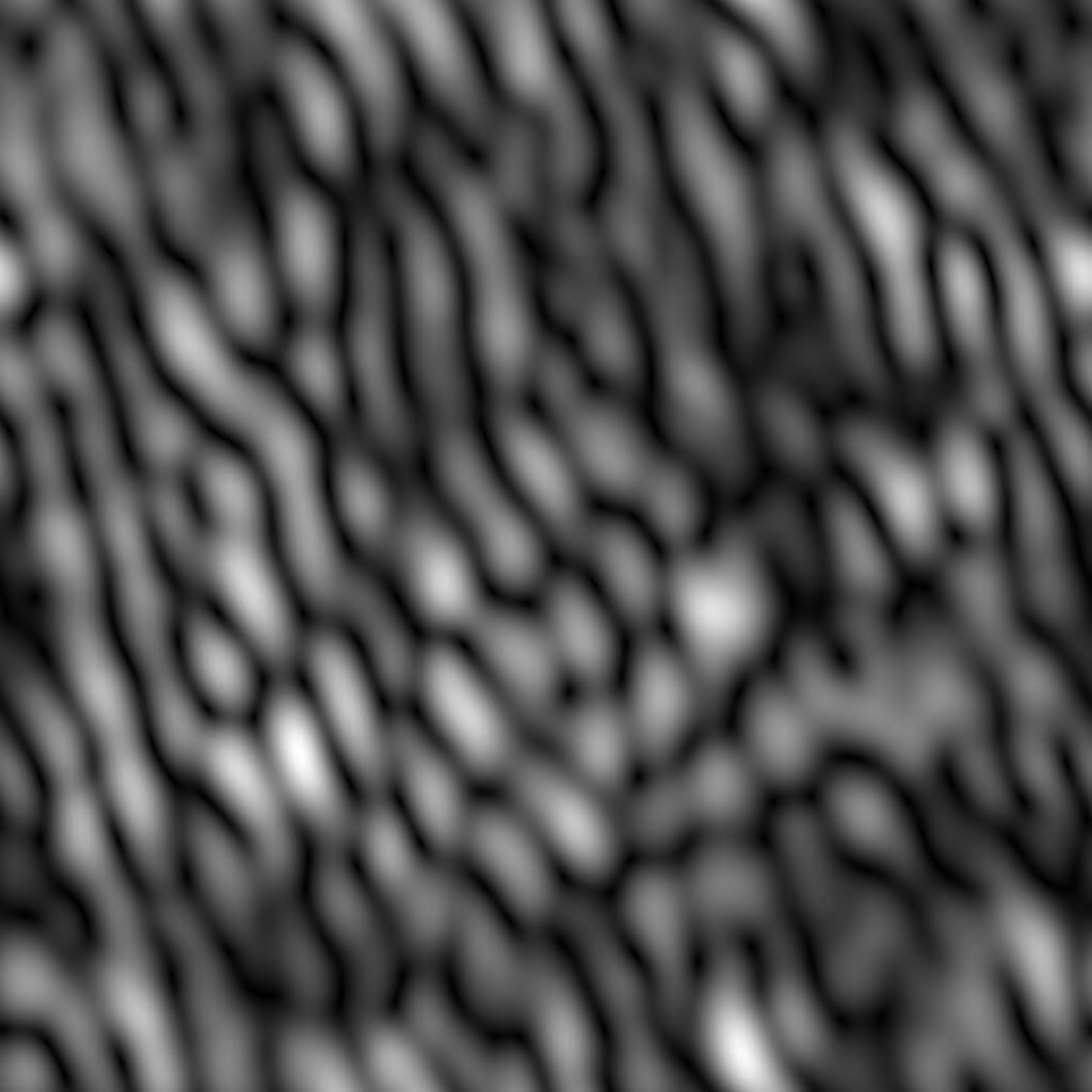}\\
\includegraphics[width=2.40cm]{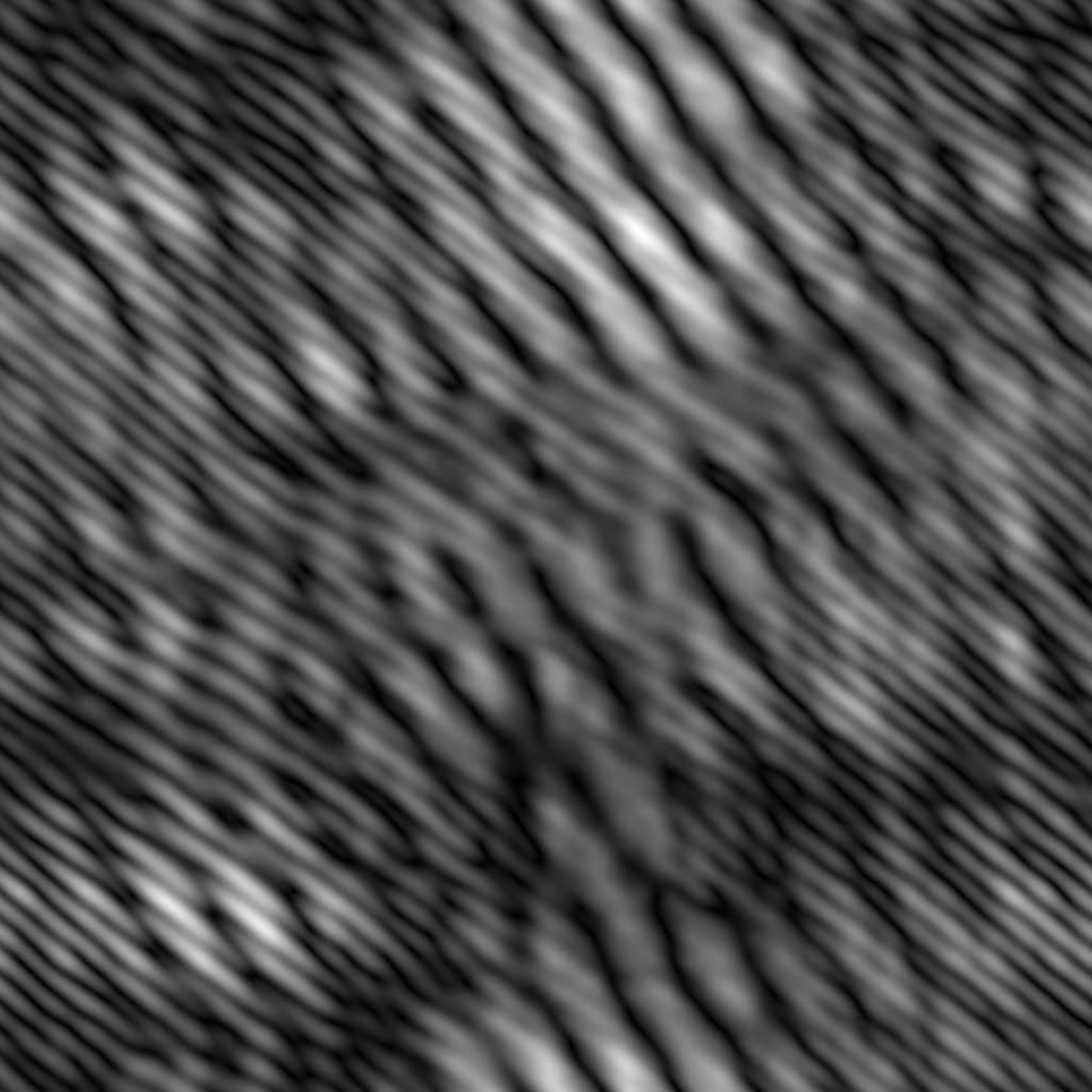}\\
\includegraphics[width=2.40cm]{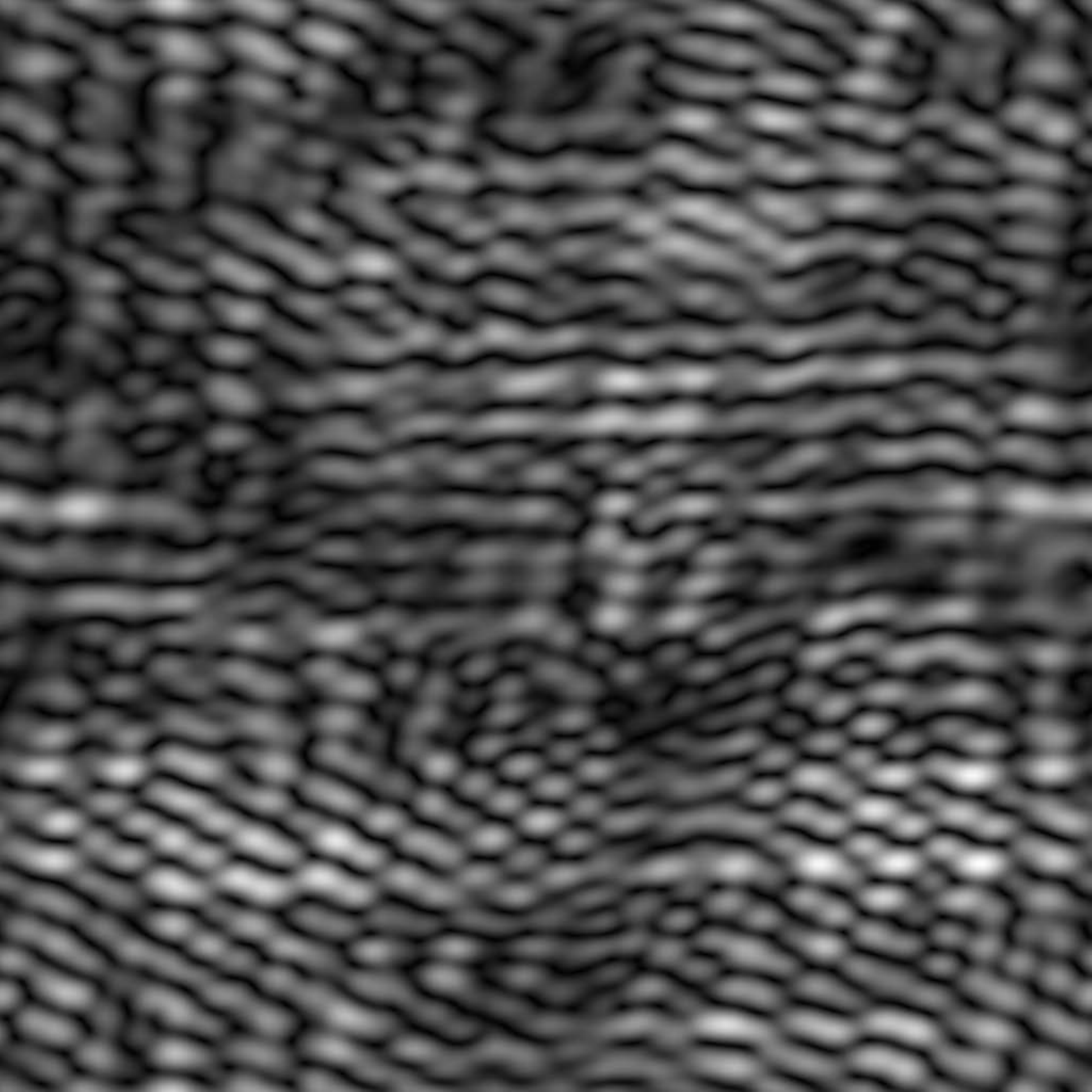}\\
\includegraphics[width=2.40cm]{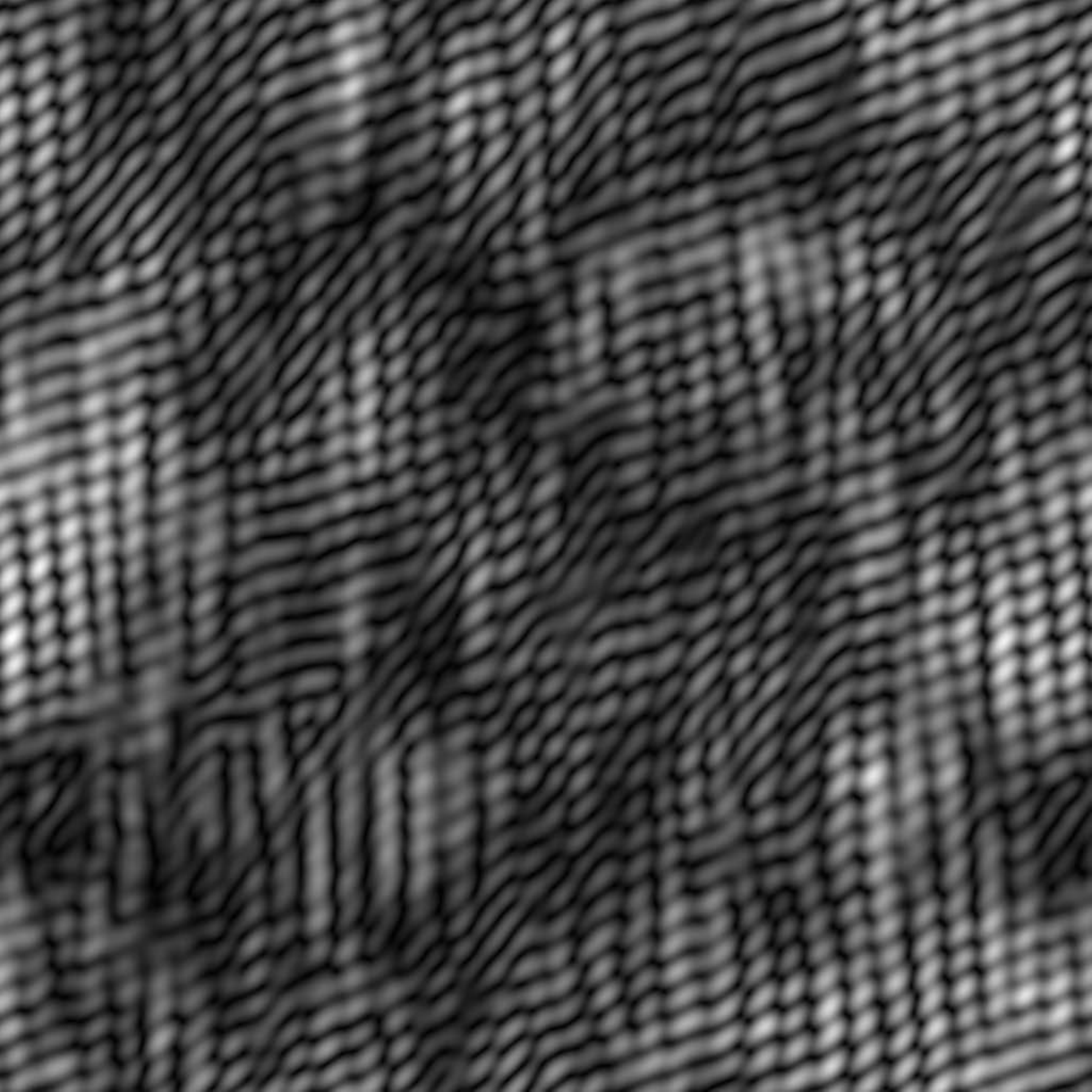}\\
\end{tabular}
\end{tabular}
\end{tabular}
}
}
&
{{
${\!\stackrel{\star}{\Rightarrow}} \:$
\begin{tabular}{@{}c@{}}
\begin{tabular}{@{}c@{}c@{}}
{} 
&
\begin{tabular}{@{}c@{}}
Class 4 \\
$\left\Vert \genfbf_{{\frak H}_{4-\bullet}} \right\Vert $\\
\includegraphics[width=2.40cm]{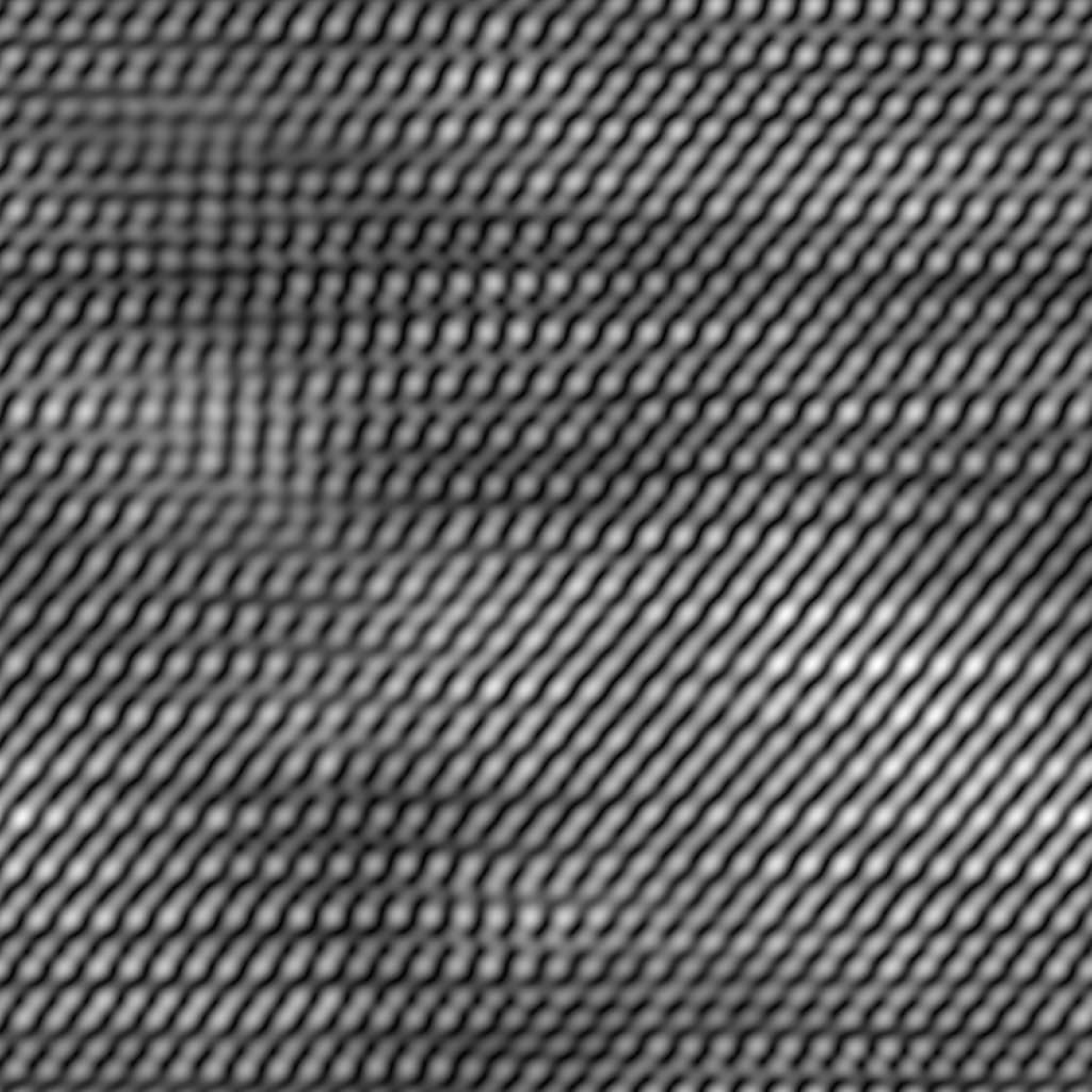}\\
\includegraphics[width=2.40cm]{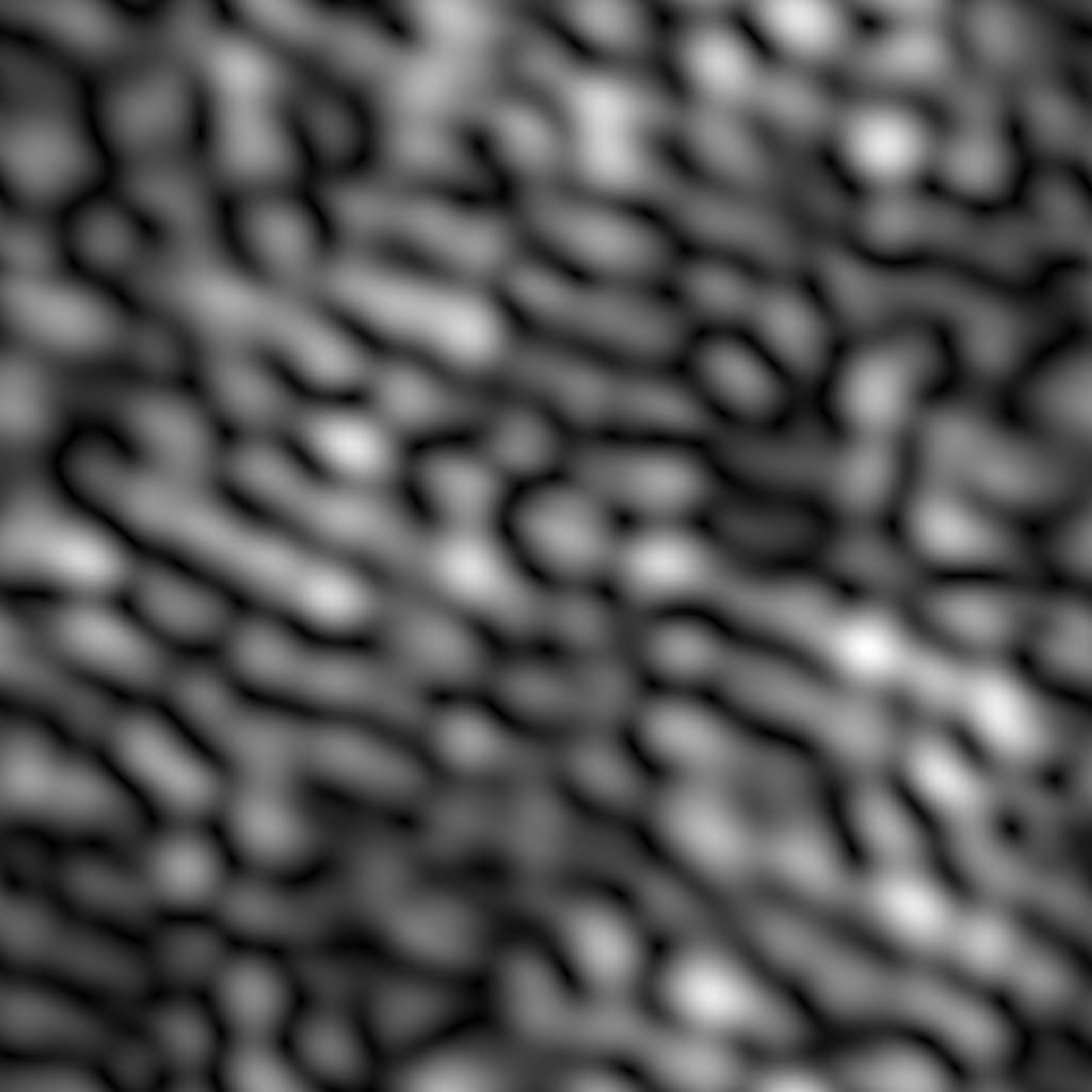}\\
\includegraphics[width=2.40cm]{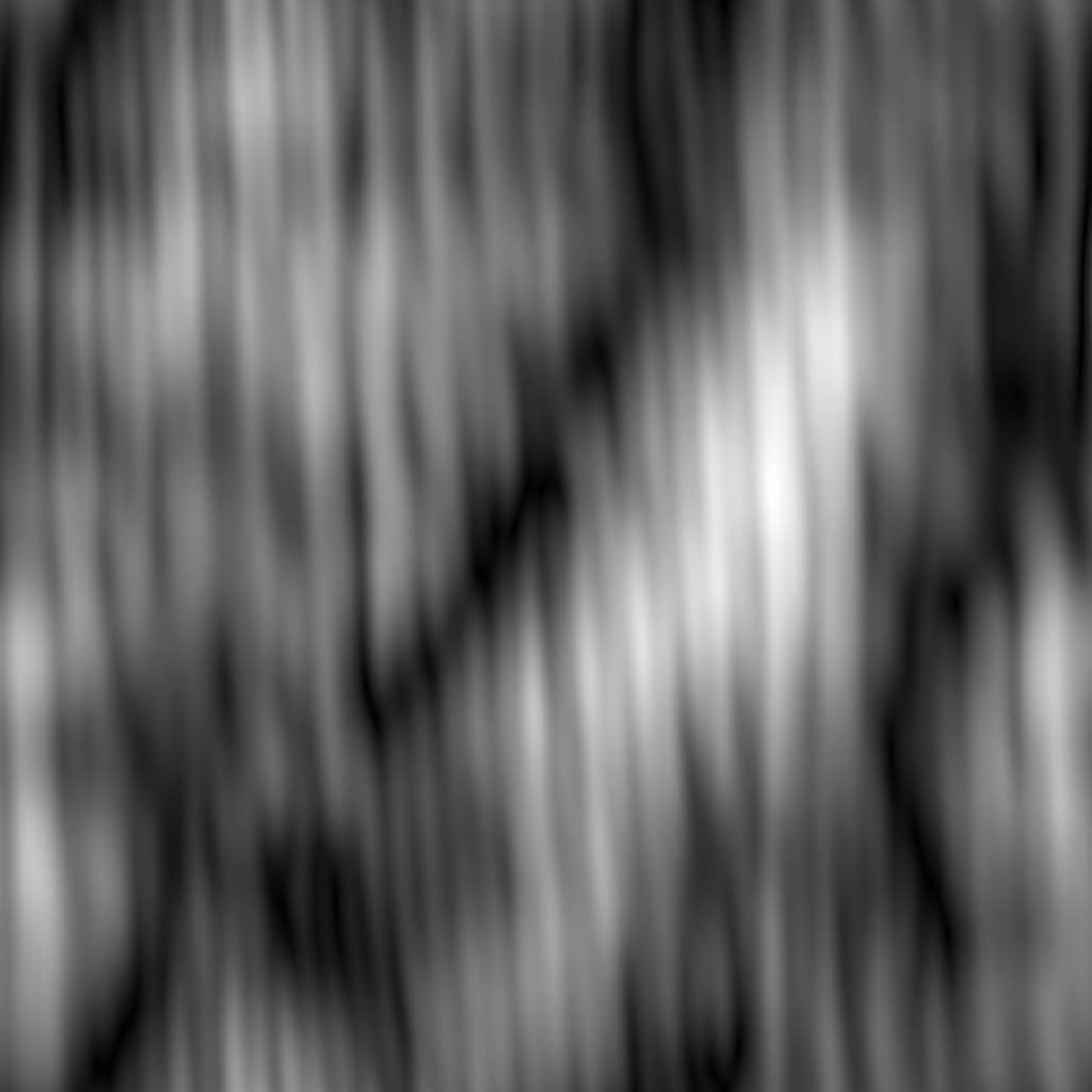}\\
\includegraphics[width=2.40cm]{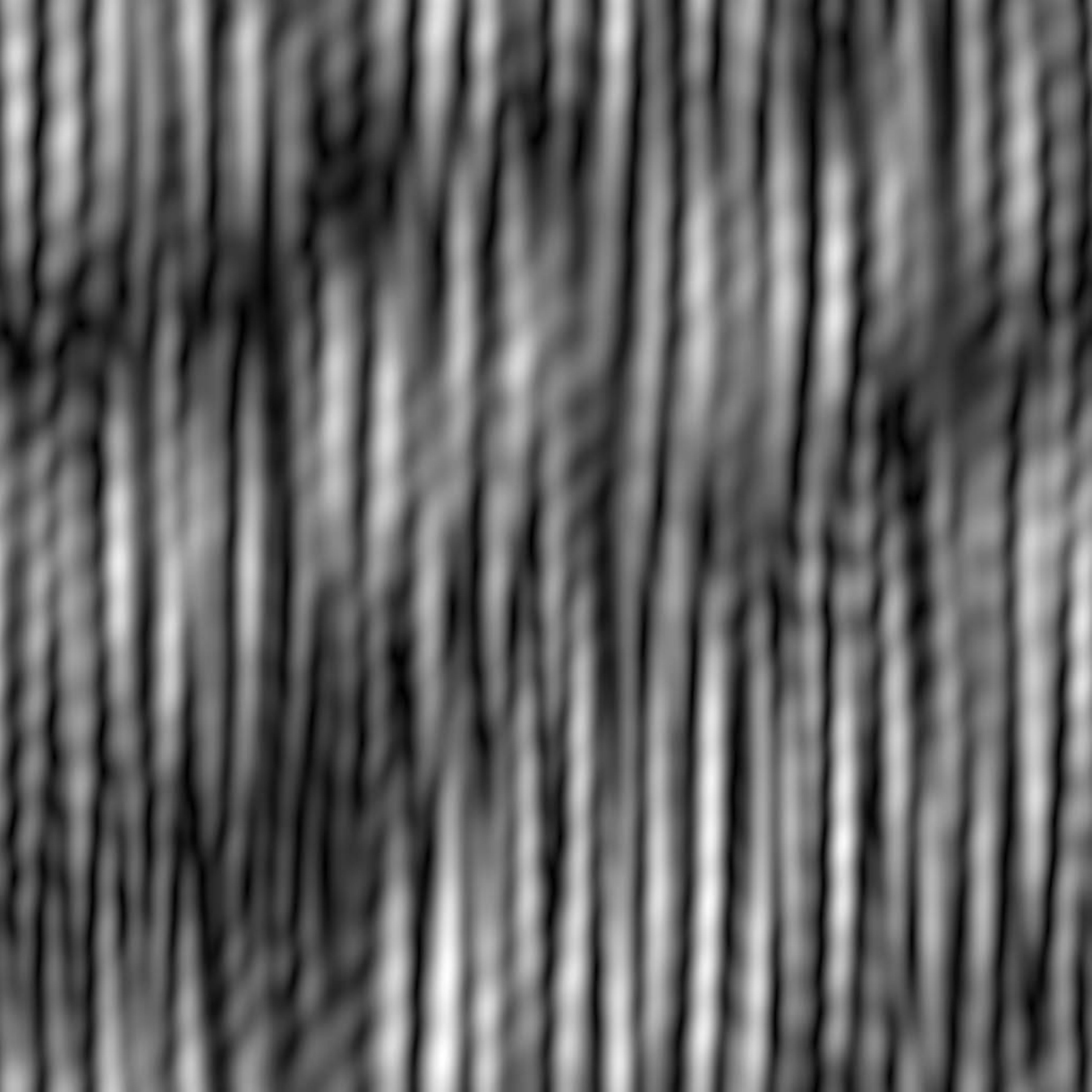}\\
\includegraphics[width=2.40cm]{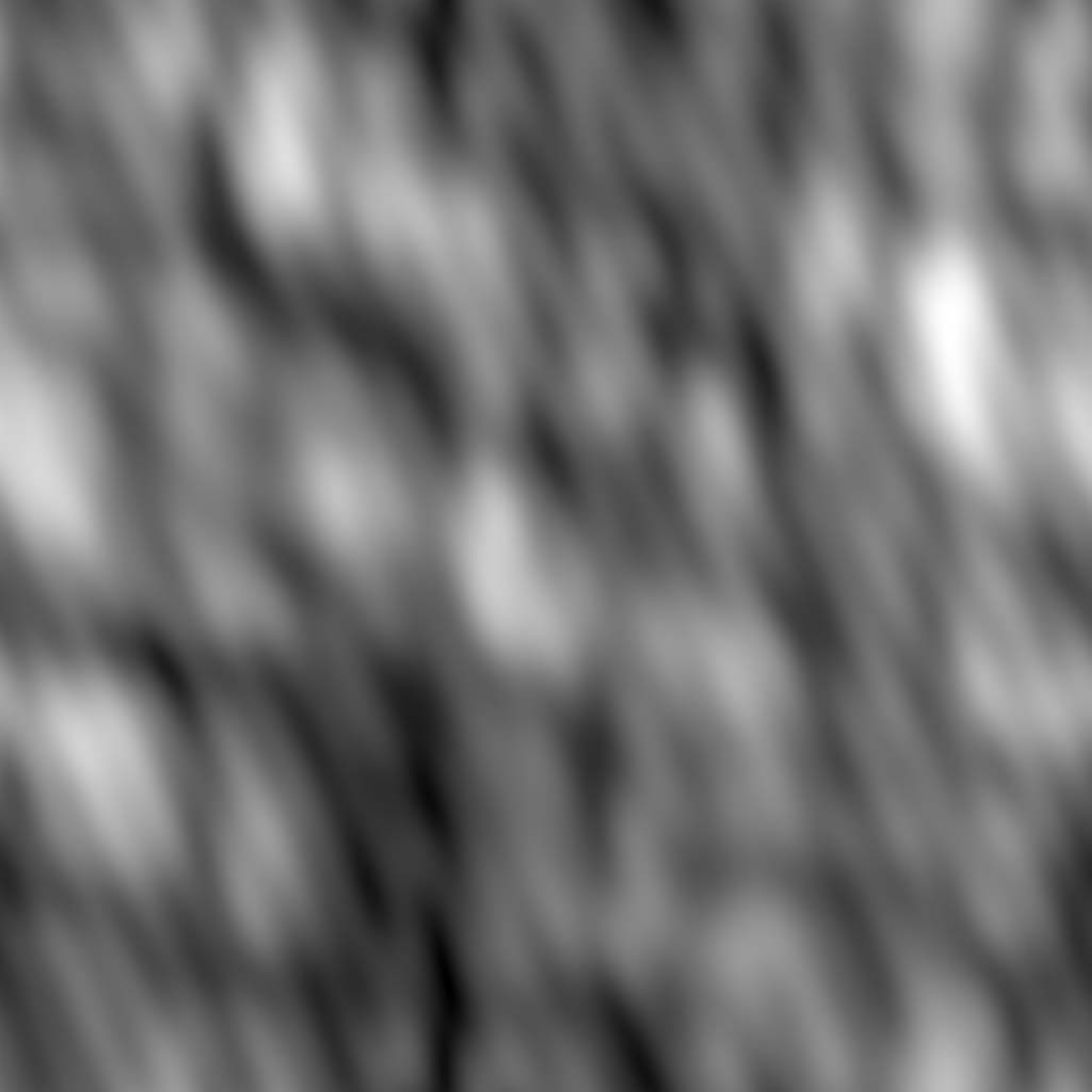}\\
\includegraphics[width=2.40cm]{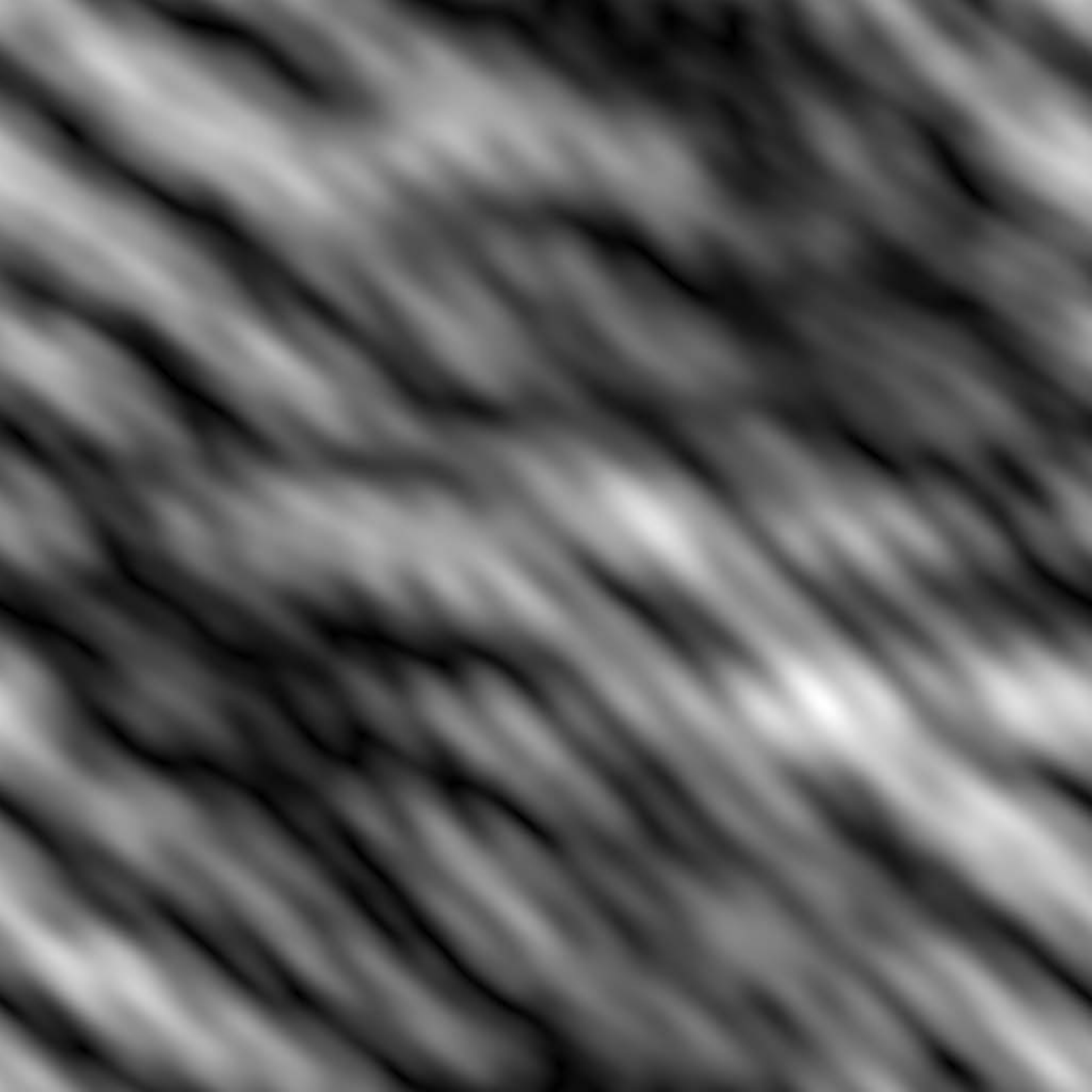}\\
\includegraphics[width=2.40cm]{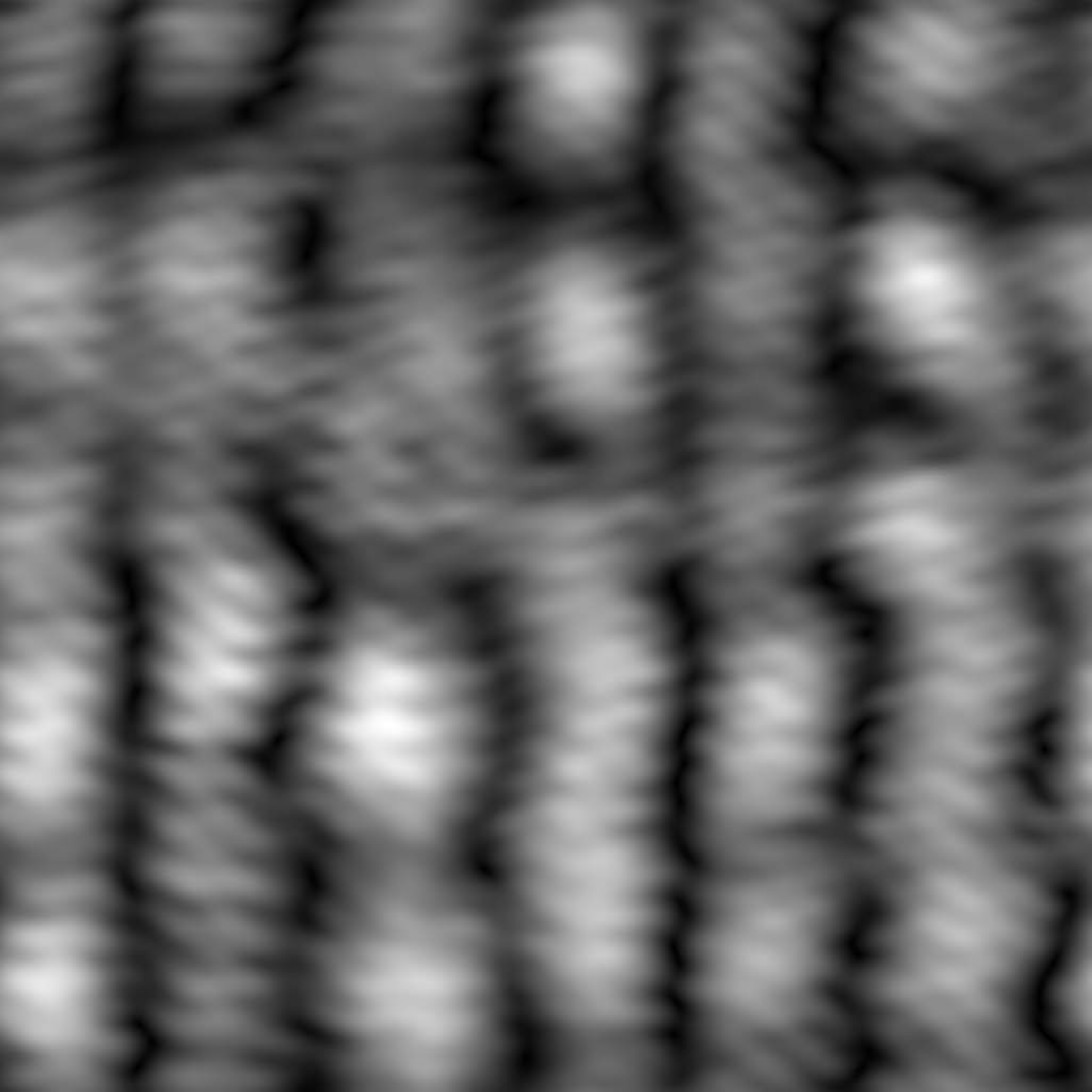}\\
\includegraphics[width=2.40cm]{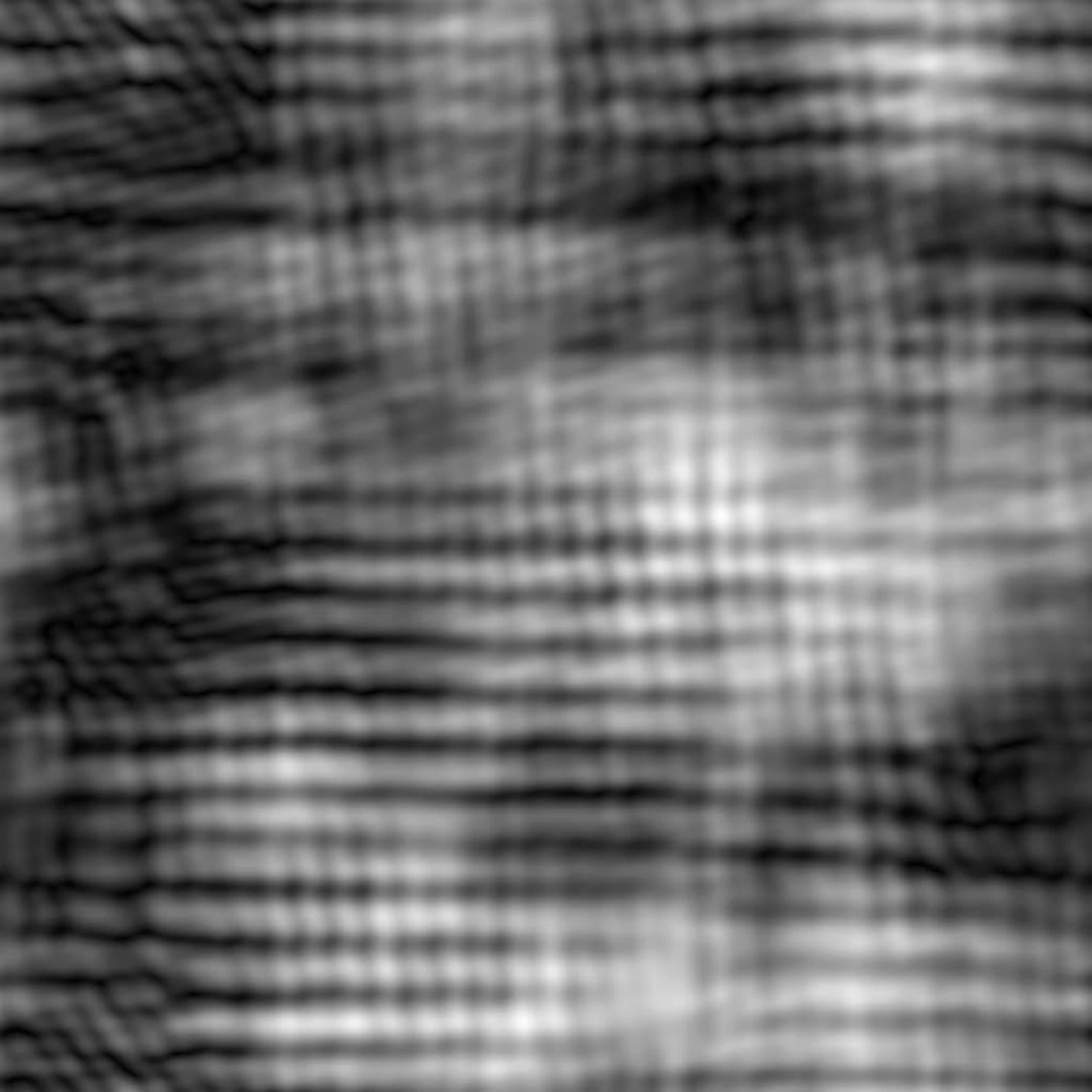}\\
\end{tabular}
\end{tabular}
\end{tabular}
}
}
\end{tabular} 
}
\caption{
Sample elements of GFBF database ${\mathcal{D}}$, where textures pertaining to any  class $Q$ of GFBF fields $\left( \genfbf_{{\frak H}_{Q-m}}\right)_m$ have been generated by using  a number $Q$ of distinct interacting modulated Fractional Brownian fields . Fractional field evolution is governed by the temporal interactions which create inter-class statistical similarities: the learning system has to made abstraction with respect to these inter-class similarities and focus on intra-class ones. 
}
\label{fig gfbf dico}
\end{figure*}

}

\subsection{
Performance validation on simulated texture series and deep learning 
}\label{sseclearn2}

The second experimental setup deployed is dedicated to performance evaluation of learning both linearities and RePSU nonlinearities in a deep CNN framework.
We will use a \underline{synthetic} database where the concept of \underline{true class} does not lead to any confusion\footnote{
Expert based labeling is far from being perfect, excepted in certain trivial contexts.
}.
This database is composed by  
Generalized Fractional Brownian Fields (GFBF, \cite{atto14icipgfbf}). GFBF is a model associated with an arbitrary number of interacting modulated fractional Brownian fields. Any of the modulated fractional Brownian fields is a long spatial memory process characterized by a given Hurst exponent and a singular spectral  point. These models make synthesis of evolution fields with rich structural content possible by using a series of spatial convolutions (linearities) and shift/modulation operators (nonlinearities).
GFBF are considered hereafter in an interaction framework where $Q$ is associated with the number of different modulated Brownian fields in interaction. 
Examples of evolution factors and 
synthesized fields are given by Figure \ref{fig gfbf dico} when the GFBF involves respectively $Q = 1, 2, 3$ and $4$  interactions.

The problem addressed is then the design of a system capable of learning the evolution factor $Q$, given an arbitrary GFBF field $X$. 
In this respect, an experimental framework has first been deployed to generate a database $\mathcal{D}$ that contains 1200 images per specified value of $Q\in\{1, 2, 3, 4\}$ (larger values of $Q$ lead to higher degrees of intricacy). A total of 4800 GFBF images has thus been generated when the number $Q$ of interacting Brownian fields pertains to the category labels $\{1, 2, 3, 4\}$, this parameter $Q$ defining the class property.
For any class, poles and Hurst parameters are generated randomly. 
An overview of the intricacy of the concept of class associated with this database is shown in Figure \ref{fig gfbf dico}, where textures pertaining to the same column pertain to the same class: this figure shows interclass dependency, a scenario that limits learning capabilities as confusion is possible between intra-class similarities (number of interactions) and inter-class similarities (remaining dependencies after field evolution). Such a challenging classification problem justifies the use of a deep CNN framework.

When using 800 textures for learning and 400 for validation per class and when learning from RePSU and ReLU based networks of Table \ref{tab cnndeep}, then the corresponding validation losses and accuracies by ten epochs are given in Table \ref{tab gfbfcnnperfo}. 
Similarly to the handwritten digit recognition results of Section \ref{sseclearn1} and in comparison with the standard CNN paradigm associated with non-learnable activations (ReLU, MISH, SWISH), the learnable activation frameworks show higher performance in general and RePSU based CNN outperforms these CNNs in terms of faster convergence to a desirable solution (increase of the validation accuracy) and the decrease validation loss.



\section{Discussion and conclusion}\label{sec conclude}

\subsection{Conclusion}\label{ssec conclusion}

In this work, we have proposed a family of nonlinear transfer functions, the RePSU functions. These functions are constructed to inherit from best qualities of ReLU and SSBS functions. RePSU based CNN involves learning nonlinear weights because parametric forms have been considered. 
The experimental results show that RePSU based CNN achieves higher performance in terms of learning and validation criteria, in comparison with ReLU, MISH and SWISH based CNNs.

\subsection{Discussion}\label{ssec discute}

We can reasonably expect to improve RePSU based CNN performance by taking more RePSU layers into account. However, computational complexity then explodes and the combination of RePSU in downstream layers and ReLU in upstream layers seems the best strategy for obtaining a good compromise for fast and efficient learning.



The main issue raised by PSWISH (defined in \cite{swish17} for the parametric form and in \cite{silu20} for the non-parametric form) is the fact that PSWISH output is not 0 even for very large negative inputs. This implies well-known limitations associated (similar to those of the sigmoid) in terms of very small but non-null gradients. 
PMISH \cite{mish20} suffers from the same default as is it non-zero almost everywhere and moreover, we have observed vanishing gradient issues during learning when $\xi$ tends to zero or is initialized close to zero. It is worth noticing that the above references have not addressed learning a series of PMISH/PSWISH $\xi$ parameters as we have done in the CNNs of Tables \ref{tab cnnbasic} and \ref{tab cnndeep}.

The RePSU-based peformance improvements reported in this paper raise an open question that relates to replacing ReLU-based deep CNN frameworks with RePSU-based shallow CNN frameworks. Since non-linearities can be handled by a single RePSU-based layer instead of using several ReLU-based, it should be indeed possible to experimentally show that few RePSU-based layers can outperform a significant amount of ReLU-based layers. By doing so, the complexity burden could be controlled/avoided, even if RePSU parameters also have to be learnt, leading to a shallow architecture whose interpretability (e.g., model size or monotonicity~\cite{intepretability}) would be better than the one of a deep framework. 
One can finally note that parameter $\beta$ is very sensitive and difficult to learn in practice: only this parameter has been set to 1 during the experiments. A specific updating strategy requiring very small gradient increments needs to be developed for learning an optimal estimate of this parameter.

\begin{center}

\end{center}

\end{document}